\documentclass[numberedappendix,english]{emulateapj}
\usepackage[T1]{fontenc}
\usepackage{babel}
\usepackage{apjfonts}

\providecommand{\LyX}{L\kern-.1667em\lower.25em\hbox{Y}\kern-.125emX\@}
\newcommand{\noun}[1]{\textsc{#1}}

\def\zperrH01{0.02}

\def\mMo{30.7}   
\def\errmMo{0.25} 

\def\DMpc{13.8}   
\def\errDMpc{1.7} 

\def\wfside{15} 
\def\pcside{10} 

\def\massyoung{(1.9\pm0.2)\times 10^5}  
\def\mYoung{      2        \times 10^5} 

\def\mOld{        7 \times 10^7} 
\def\massold{   6.7\times  10^7} 

\def\kms{~\rm km~s^{-1}}

\begin{document}

\title{The Stellar Content of the Southern Tail
of NGC 4038/9 and a Revised Distance
\altaffilmark {1}}

\altaffiltext{1}{ Based on observations made with the NASA/ESA Hubble
Space Telescope obtained at the Space Telescope Science Institute,
which is operated by the Association of Universities for Research
in Astronomy, Inc., under NASA Contract NAS-5-2655.  These observations
were made in connection with proposal GO-6669.}

\author{Ivo Saviane}

\affil{European Southern Observatory, 
3107 Alonso de Cordova,  Vitacura -- Casilla 19001, Santiago 19, Chile
and Department of Physics \& Astronomy, University of California at Los 
Angeles, Math-Sciences 8979, Los Angeles, CA 90095-1562}

\email{isaviane@eso.org}

\author{J. E. Hibbard}

\affil{
National Radio Astronomy Observatory,1 520 Edgemont Road, Charlottesville, 
VA 22903 
}

\email{jhibbard@nrao.edu}

\and{}

\author{R. Michael Rich}

\affil{
Department of Physics \& Astronomy, University of California at Los Angeles,
Math-Sciences 8979, Los Angeles, CA 90095-1562
}

\email{rmr@astro.ucla.edu}

\begin{abstract}

We have used the {\it Hubble Space Telescope} and Wide Field Planetary
Camera~2 to image the putative tidal dwarf galaxy located at the tip
of the Southern tidal tail of NGC 4038/9, the Antennae.  We resolve
individual stars, and identify two stellar populations.  Hundreds of
massive stars are present, concentrated into tight OB associations on
scales of $200$~pc, with ages ranging from $2$-$100$~Myr.  An older
stellar population is distributed roughly following the outer contours
of the neutral hydrogen in the tidal tail; we associate these stars
with material ejected from the outer disks of the two spirals.  The
older stellar population has a red giant branch tip at $I=26.5\pm 0.2$
from which we derive a distance modulus $(m-M)_0=\mMo\pm\errmMo$.  The
implied distance of $\DMpc\pm\errDMpc$ Mpc is 
significantly smaller
than 
commonly quoted distances for NGC 4038/9. 
In contrast to the previously studied core of the merger, we find no
super star clusters.  One might conclude that SSCs require the higher
pressures found in the central regions in order to form, while
spontaneous star formation in the tail produces the kind of O-B star
associations seen in dwarf irregular galaxies.  The youngest
population 
in the putative tidal dwarf
has a total stellar mass of $\approx \mYoung M_\odot$,
while the old population has a stellar mass of $\approx 
\mOld ~M_\odot$.  
If our smaller distance modulus is correct, it has
far-reaching consequences for this proto-typical merger. Specifically,
the luminous to dynamical mass limits for the tidal dwarf candidates
are significantly less than $1$, the central super star clusters have
sizes typical of galactic globular clusters rather than being 
$1.5$ times as large,
and the unusually luminous X-ray population becomes both less
luminous and less populous.

\end{abstract}


\section{Introduction  } \label{sec:intro}

Interactions and mergers of galaxies are
becoming increasingly recognized 
as perhaps the most important physical process affecting
galaxy evolution.  In the context of hierarchical clustering models of
galaxy formation, luminous galaxies such as the Milky Way could have
been formed as the product of early mergers of less massive galaxies
(Kauffmann \& White \citeyear{kauffmannWhite93}).  Considering the
crucial role that galaxy merging plays in the evolution of massive
galaxies in CDM cosmology, it is important to consider also the
derivative effects that this merger activity has on the evolution of
galaxy populations.

Since the pioneering work of Toomre \& Toomre (\citeyear{TT}) it has
been known that dramatic tidal tails are among the most striking signs
of a merger in progress.  It has been suggested that the material in
tidal tails might detach from the parent galaxies and collapse into
gravitationally bound clumps, leading to independent objects (so-called
Tidal Dwarf Galaxies or TDGs) that might form new stars and be later
recognized as dwarf galaxies (Zwicky \citeyear{zwicky56}; Schweizer
\citeyear{schweizer78}, hereafter S78; Barnes \& Hernquist
\citeyear{barnes-hernquist92}; Duc et al. \citeyear{duc_etal97},
Duc \& Mirabel \citeyear{duc_mirabel98}).  This process could be more
significant at high redshift where the merger rate is dramatically
higher.

\begin{figure*}
  \plotone{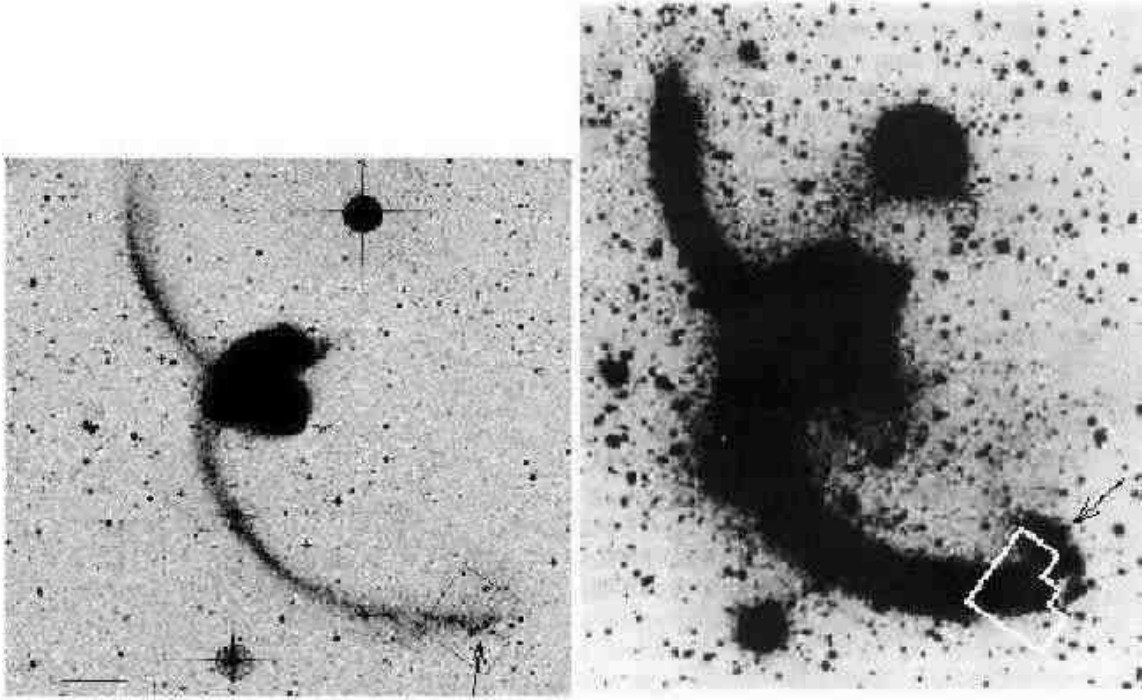}
\caption{{\em Left panel}:
The position of the \noun{wfpc2} field is represented by the black
outline over a reproduction from the DSS field (North is up and East
to the left).  The bar at the lower left of the figure is \protect\(
2\protect \) arcmin, and the \noun{wfpc2} field is \protect\( \sim
10\protect \)~arcmin SW from the nuclei of the two galaxies. NGC~4038 is
the northern object (plus the southern tidal tail), and NGC~4039 is the
southern one (plus the northern tidal tail). The arrow points to the
object identified by MDL92 as a TDG. {\em Right panel}: Reproduction of
the CTIO plate in Schweizer (\citeyear{schweizer78}), where the arrow
points to the low surface brightness object he identified as a candidate
TDG beyond the tip of the tail. The white outline represents the
\noun{wfpc2} field.
\label{fig:dss} }
\end{figure*}

One of the best cases for a TDG actually in the process of formation
is near the end of the long, curving southern tidal tail of the well
known nearby merging system NGC 4038/9, ``The Antennae''.  Schweizer
(\citeyear{schweizer78}) was  the first to discuss the tip of
the Southern tail in some detail. Making use of stacked photographic
images obtained using the CTIO 4-m telescope (reproduced in
Fig.~\ref{fig:dss}), Schweizer noted the presence of a low surface
brightness object (\( V>25 \) magnitudes~arcsec\( ^{-2} \)) slightly
\emph{beyond} the end of the tail (indicated by an arrow in the right
panel of Fig.~\ref{fig:dss}).  He noted that at the adopted distance
of $\sim 28$~Mpc, this feature would have an absolute magnitude \(
M_{V}\simeq -16.5 \) making it roughly twice as luminous as
the local group dwarf irregular galaxy IC~1613. Furthermore, the
then-current WSRT H{\sc i} maps of this system by van der Hulst (1979)
showed a strong concentration of atomic gas coincident with this
extension, and Schweizer hypothesized that it was a distinct dwarf
stellar system, possibly created during the interaction as envisioned
by Zwicky (1956).

Besides this possible TDG, S78 observed star formation taking place 
\emph{within} the tip of the tail (indicated by an arrow in the left 
panel of Fig.~\ref{fig:dss}).  This region of the tail was
subsequently studied by Mirabel et al. (\citeyear{mirabel}; hereafter
MDL92). MDL92 described a chain of H{\sc ii} regions delineating a
twisting stellar bar embedded in an envelope of diffuse optical
emission.  Optical spectroscopy of three of the H{\sc ii} regions
(denoted as Regions I, II, and III by MDL92) showed them to be ionized
by stars as young as 2 Myr (Region I) and 6 Myr (Regions II and III),
and to have an oxygen abundance of \( 12+\log (\rm O/H)=8.4 \)
(i.e. $\simeq 1/3$ solar). The total luminosity of the three H{\sc ii}
regions was found to be \( \sum L(H\alpha ) = 3.9\times
10^{39}\)~erg\, s\(^{-1}\), and the apparent magnitude of the entire
region is \( V=17.3\pm 0.3 \), corresponding to \( M_{V}=-15.5 \)
(all values referenced to their assumed distance of \( 33.2 \) Mpc,
i.e. \( m-M=32.6 \)). These properties, along with the atomic gas mass
of \( M(HI)= 10^{9} \) \( M_{\odot } \) and derived dynamical mass of
\( 8.3\times 10^{9}\, M_{\odot } \), are similar to those of dwarf
irregular galaxies, and MDL92 concluded that this regions was in fact
a dwarf irregular galaxy forming out of the tidal debris.

More recent radio observations have mapped the total extent of the
H{\sc i} within the tidal tails (Gordon et al. \citeyear{gordon_etal01}, 
Hibbard et al. \citeyear{hibbard_etal01}). These observations show that, 
while the density of the atomic gas is enhanced in the vicinity of the
putative TDGs identified by S78 and MDL92, the entire tail is a
continuous gas rich structure. The gas kinematics suggest that the
tail bends back along our line of sight just in the vicinity of the
TDG candidates, giving rise to the appearance of a distinct clump of
gas associated with the putative tidal dwarfs. No clear kinematic
signature of a self-gravitating entity of the 
mass
suggested by MDL92
was seen, but the geometric velocity gradients caused by the tail
bending away from our line of sight may mask the kinematic signature
of smaller self-gravitating condensations (Hibbard et al.
\citeyear{hibbard_etal01}). Even though the dynamical nature of
the TDG candidates is still in question, the H{\sc i} observations
show that the atomic gas in the vicinity of star forming regions I,
II, and III of MDL92 is denser than anywhere else in the system.

To study further the star forming regions identified by S78 and
studied by MDL92, and to examine the distribution and ages of the
underlying old and young stellar populations of the TDG candidate
identified by MDL92, we targeted this region for broadband UBVI
observations with Wide Field and Planetary Camera~2 (\noun{wfpc2}) 
on board the Hubble Space Telescope (HST). Figure~\ref{fig:dss} 
illustrates the location of our single \noun{wfpc2} field relative 
to NGC 4038/9. In addition to investigating the population of a TDG
candidate, these observations also afford us the chance to investigate
star formation taking place in an environment relatively isolated from
the large scale structure of a disk, in the far outskirts of a
galactic potential well. We stress that a study of the S78 object with
the recently installed ACS would be valuable as well, as the
population of this object may well sample some of the original disk
population of NGC 4038 and additionally give an improved red giant
branch tip distance test of our proposed short modulus.

\begin{figure*}
  \plotone{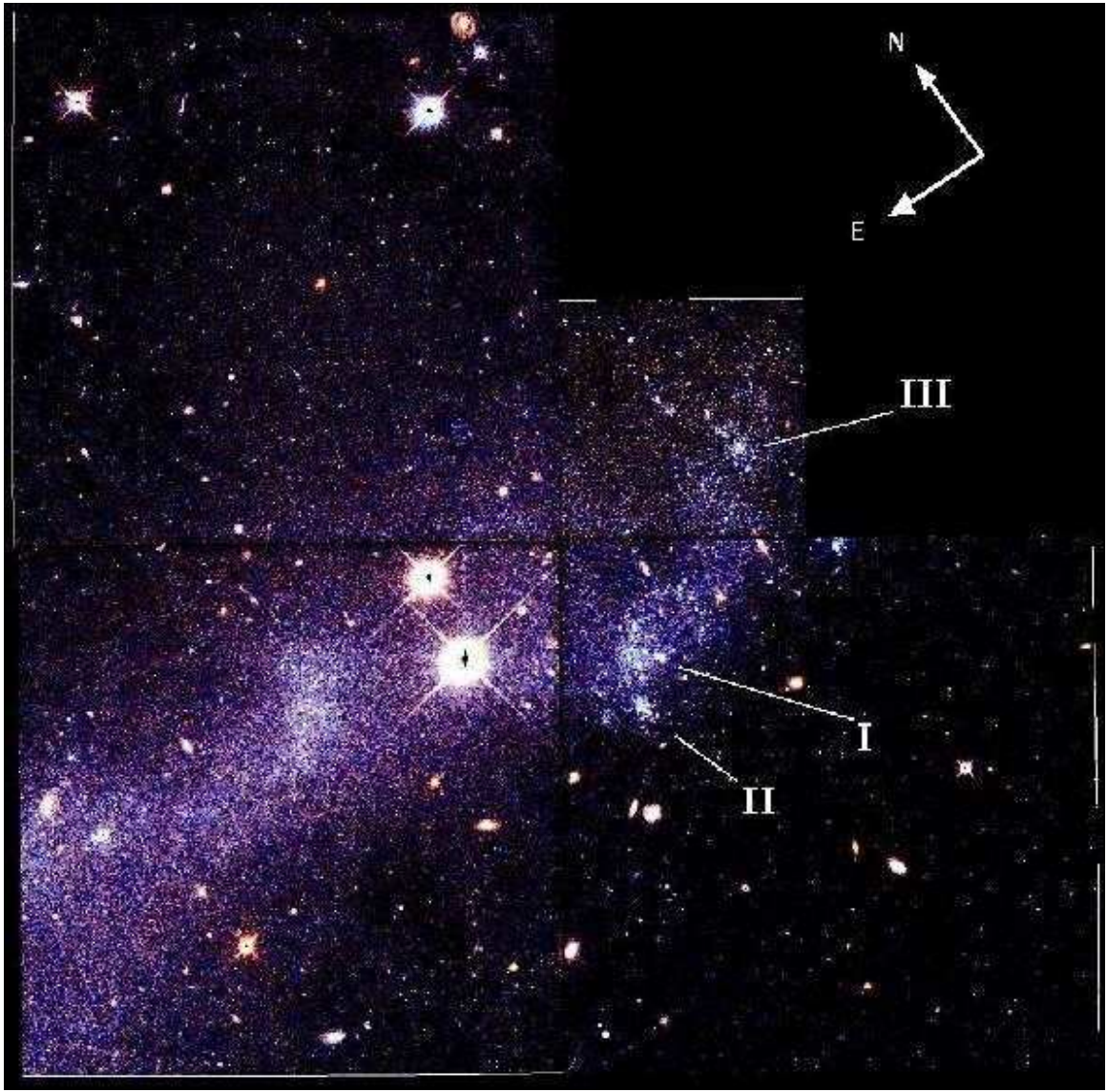}

\caption{True-color image of our HST field, 
where red has been assigned to the F814W image, blue to the F555W, and
green to an image made from a combination of both filters. The extent
and orientation of the field is outlined in Fig.~\ref{fig:dss}. The 
three star formation regions studied by Mirabel et al. 
(\citeyear{mirabel}) are marked along the bar-like structure that 
extends westward from the two bright stars. 
\label{fig:hst-mosaic}}
\end{figure*}

A color image of the HST mosaic is shown in Fig.~\ref{fig:hst-mosaic},
and a visual inspection of the image gives a clear impression of the
population, even without the formality of a color-magnitude diagram (CMD).
The light of the tidal tail region imaged is dominated by bright blue
stars indicative of a young stellar population (which contrast
strongly with the reddish background galaxies).  Fainter and redder
stars indicative of an older population can also be seen.  The
youngest and brightest blue stars are located along a bar-like
structure that extends from the two bright foreground stars in the
upper right corner of the WF3, crosses the upper left corner of the
WF4, and ends in the lower right corner of the PC. These stars form
groups that look like OB associations, and the three most prominent
ones coincide with the location of the three H{\sc ii} regions studied
by MDL92. Following that study, we refer to these regions as Regions
{\sc I, II, III}, and have labeled them as such in 
Fig.~\ref{fig:hst-mosaic}. Besides these active sites of star
formation, another concentration of young stars can
be seen in the WF3 chip, East of the two foreground stars.

The paper is organized as follows. The data reduction and calibration
are presented in \S\ref{sec:obs-reduc}, and the resulting CMD is used
in \S\ref{sec:distance} to find the distance to the Antennae. We
interpret the CMD with the aid of that of NGC~625 (a star forming dwarf
irregular galaxy) and theoretical isochrones, and the distance modulus
is then found using the luminosity of the tip of the red giant branch
of the old population. The stellar content of the NGC~4038 tail is
discussed in \S\ref{sec:star-content}, and in particular, we discuss
the spatial distribution of populations as a function of age in
\S\ref{sec:space-age}. A number of stellar associations are defined,
and their morphology is compared to the central star clusters in
\S\ref{sec:recent-sf}. Their CMD is interpreted in \S\ref{sec:comp-isos}, 
and an estimate of the mass in young and old stars is obtained in 
\S\ref{sec:youngstars} (plus Appendix~\ref{sec:estimating_young}) 
and \S\ref{sec:fornax}, respectively.  In order to interpret the
luminosity function of the young population, a model of constant and
continuous star formation is developed in Appendix~\ref{sec:simple-sf},
and the luminosity function of the old population is compared to that of
a Virgo dwarf elliptical in Appendix~\ref{sec:vcc1104}. Our conclusions
are given in
\S\ref{sec:discussion}.

\section{Observations, reductions and calibrations} \label{sec:obs-reduc}

The field centered at $ (\alpha ,\delta ,\rm epoch) =
$~($12$:$01$:$25.6$, $-19$:$00$:$31.9$, J2000) was imaged on November
23 and 24 1998, and January 4 and 5 1999. The outline of the HST
\noun{wfpc2} field of view is indicated on a Digital Sky Survey (DSS)
image of NGC~4038/9 in Fig.~\ref{fig:dss}.  Four filters were
employed, for a total observing time of 1.4h in F336W, 0.7h in F450W,
2.1h in F555W and 2.9h in F814W.  The exposure times for the two
shortest wavelength filters were not long enough to allow the
detection of stars, so our emphasis will be on the F555W and F814W
frames.


\begin{figure}
  \plotone{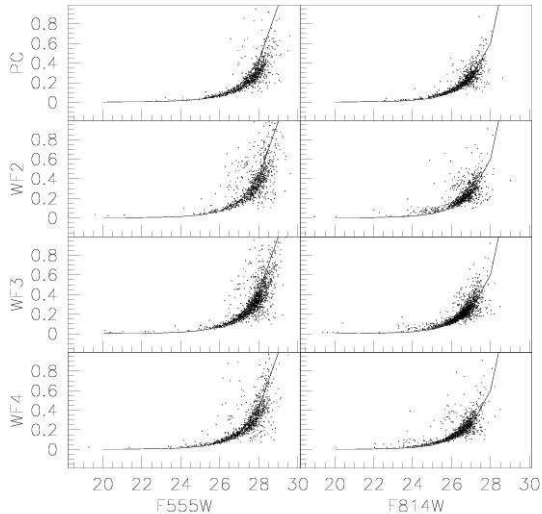}
\caption{\noun{allframe} internal photometric errors. 
Absolute errors (in magnitudes) for the two filters are plotted 
in the left ($F555W$ filter) and right ($F814W$ filter) columns 
vs. the apparent magnitude.  The solid curves represent the 
expected error based on the \noun{wfpc2} exposure time calculator. 
\label{fig:photerrors}}
\end{figure}


\subsection{Reductions}

Our procedures for data reduction and calibration are based on
the \noun{daophot~II/allframe} package (Stetson \citeyear{stetson87},
\citeyear{stetson94}). The procedures have been thoroughly described 
in Piotto et al. (\citeyear{piotto_etal02}) and Saviane et
al. (\citeyear{saviane_etal02}), so only the general scheme is
recalled here, which involves (a) image pre-processing, (b)
profile-fitting photometry within a small radius, (c) correction for
time-dependent charge-transfer efficiency (CTE), (d) growth-curve
analysis and correction to total magnitudes, (e) calibration to the
standard system, and finally (f) astrometry.

The images are prepared for the photometry by masking out vignetted
pixels, and bad pixels and columns.  The coordinates of detected
sources are measured from a master frame, combining all of the images
in all filters.  These positions are used by \noun{allframe}, which
performs the photometry by fitting a point-spread-function (PSF)
simultaneously to all stars of the individual frames. Since there are
not sufficiently bright and isolated stars in any of the four chips,
we are not able to construct the PSFs.  Instead, we use the models
extracted by P.B. Stetson (private communication) from a large set of
uncrowded and unsaturated HST \noun{wfpc2} images.  The resulting
F555W and F814W magnitude lists are combined to create a raw
color-magnitude (CMD) diagram.

\subsection{Calibration}

\begin{deluxetable*}{lll lll lll l}
\tablewidth{0pc}
\tablecaption{Photometry of Individual Stars. \label{tab:elec-catalog}}
\tablehead{
\colhead{ID}        & \colhead{RA               }  & \colhead{DEC}  &
\colhead{$V_0$}     & \colhead{\small $F555W_0$        }  & \colhead{$\epsilon_{F555W}$ } &
\colhead{\small $(V-I)_0$} & \colhead{$C_0^{\rm HST}$}  &
\colhead{\small $\epsilon_{C^{\rm HST}}$} & 
\colhead{\#}}
\startdata
     1 &  12:01:28.625 & -18:59:06.43 &    18.219 &    18.244 &     0.021 &     0.880 &     0.870 &     0.027  &     2\\
     2 &  12:01:25.333 & -19:01:34.86 &    19.277 &    19.302 &     0.023 &     1.102 &     1.089 &     0.027  &     4\\
     3 &  12:01:25.106 & -18:59:33.80 &    19.617 &    19.639 &     0.035 &     0.675 &     0.667 &     0.039  &     2\\
     4 &  12:01:32.635 & -19:00:32.12 &    20.102 &    20.126 &     0.035 &     0.717 &     0.709 &     0.037  &     3\\
     5 &  12:01:25.439 & -18:59:44.37 &    20.276 &    20.298 &     0.023 &     1.306 &     1.290 &     0.026  &     2\\
     6 &  12:01:32.127 & -19:00:56.46 &    20.372 &    20.298 &     0.015 &     2.875 &     2.830 &     0.141  &     3\\
     7 &  12:01:28.752 & -19:00:41.55 &    21.036 &    21.058 &     0.020 &     0.639 &     0.632 &     0.110  &     3\\
     8 &  12:01:29.779 & -19:00:08.29 &    21.092 &    21.117 &     0.029 &     1.017 &     1.004 &     0.034  &     2\\
     9 &  12:01:27.784 & -19:01:30.55 &    21.360 &    21.382 &     0.020 &     1.290 &     1.274 &     0.027  &     4\\
    10 &  12:01:32.175 & -19:00:46.49 &    22.170 &    22.151 &     0.023 &     2.248 &     2.215 &     0.027  &     3\\
    11 &  12:01:30.112 & -19:01:20.58 &    22.355 &    22.338 &     0.026 &     2.210 &     2.178 &     0.030  &     3\\
    12 &  12:01:29.672 & -19:00:53.06 &    22.617 &    22.530 &     0.026 &     3.000 &     2.953 &     0.030  &     3\\
    13 &  12:01:25.546 & -19:00:18.67 &    22.747 &    22.756 &     0.019 &     0.198 &     0.196 &     0.026  &     1\\
    14 &  12:01:25.275 & -19:00:39.61 &    23.209 &    23.230 &     0.048 &     0.569 &     0.563 &     0.063  &     1\\
    15 &  12:01:25.086 & -19:00:58.72 &    23.367 &    23.380 &     0.029 &     0.311 &     0.307 &     0.035  &     4\\
    16 &  12:01:24.941 & -19:00:41.77 &    23.392 &    23.416 &     0.021 &     0.797 &     0.788 &     0.028  &     1\\
    17 &  12:01:27.231 & -19:00:55.47 &    23.462 &    23.468 &     0.020 &     0.132 &     0.131 &     0.032  &     4\\
    18 &  12:01:28.399 & -19:01:23.96 &    23.505 &    23.529 &     0.029 &     0.832 &     0.823 &     0.035  &     4\\
    19 &  12:01:25.583 & -19:00:29.44 &    23.686 &    23.699 &     0.027 &     0.293 &     0.290 &     0.035  &     1\\
\enddata
\tablecomments{The coordinates are J2000,
$C_0^{\rm HST}=(F555W-F814W)_0$, and \# is the \noun{wfpc2} chip number, with
1~=~PC, 2~=~WF2, etc. The
complete version of this table is in the electronic edition of the
Journal.  The printed edition contains only a sample.}
\end{deluxetable*}

In the next step, the photometry is calibrated to the standard system,
following Dolphin (\citeyear{dolphin-cte}; D00), and assuming an average
reddening in front of NGC~4038/9 of \( E_{B-V}=0.046 \) (Schlegel et al. 
\citeyear{schlegel_etal98}). This procedure accounts for both the 
time/counts dependence of the CTE and the variation of the effective 
pixel area across the \noun{wfpc2} field of view, and it yields final 
calibrated magnitudes in  the 
Johnson--Kron--Cousins
and the HST photometric 
systems.

The aperture corrections (to the reference aperture of \( 0\farcs5 \)
of Holtzman et al. \citeyear{holtzman}; H95) could not be computed in
the usual way, since we lacked the necessary bright isolated objects.
Thus, we followed an approach closely resembling that of Ibata et
al. (\citeyear{ibata-m4}). Artificial images were created using
the \noun{iraf/mknoise} task, which reproduced the observed
background+noise of our reference scientific frames (one frame per
filter and per \noun{wfpc2} chip). On this background, the
\noun{daophot/add} routine was used to add a number of stars
spanning an instrumental magnitude range from \( 14 \) to \( 20 \),
making use of the available PSFs. 
For the details of this process, see Saviane et
al. (\citeyear{ivo-fornax}). These artificial images were finally used
to compute the aperture corrections, with uncertainties of $\pm
0.05$~magnitudes.

\begin{figure}
  \plotone{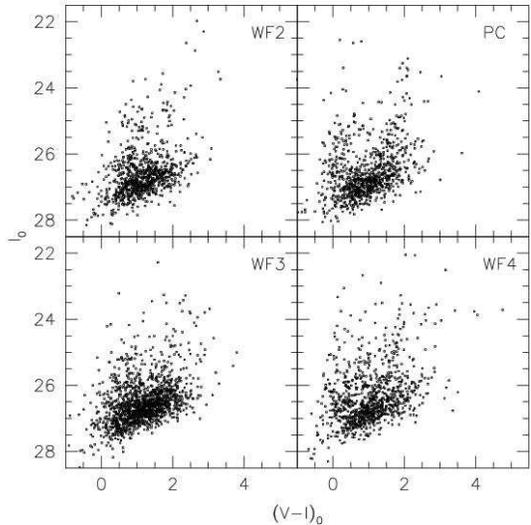} 
\caption{The panels in this figure show
the color magnitude diagram of each of the four \noun{wfpc2} chips, according
to the labels in the upper-right corner. True $V-I$ colors and $I$
magnitudes are displayed along the abscissae and ordinates,
respectively.
\label{fig:cal4} }
\end{figure}

The J2000 positions of our objects were found through the
\noun{metric} task in the \noun{iraf/stsdas} package. After
processing, the world coordinates were appended to the photometry, and
finally a single catalog was obtained by merging the four chip
measurements. This catalog forms the basis of our discussion of the
stellar content of this region of NGC~4038. We estimate a total
systematic error of $\pm 0.05$~magnitudes, dominated by the
uncertainty on the aperture corrections.

\subsection{Characterization of the photometry}

Figure~\ref{fig:photerrors} shows the resulting photometric errors as
returned by \noun{allframe} for both filters, and for each of the
\noun{wfpc2} chips. One can immediately see that most of the stars are
fainter than the $24^{\rm th}$ magnitude, and that below $F555W=28$,
or $F814=27$, the incompleteness of the data starts to be severe. The
typical trend of rising errors with magnitude can be seen (with
smaller scatter in the PC and WF4 CCDs), and most of the stars in the
field are affected by errors of a few $0.1$ magnitudes. The solid
lines represent the expected photometric error based on the $S/N$
predicted by the \noun{wfpc2} exposure time calculator for the WF
chips and an A0 star, within our \noun{daophot} fitting radius of
2~px. The observed errors are (on average) $10\%$ smaller than
expected for F555W, and ca. $20\%$ larger than expected for F814W.  
A few lines of the final photometric catalog are reported in
Table~\ref{tab:elec-catalog}, where the columns contain, from left to
right, an identifier (ID), the J2000 equatorial coordinates (RA and
DEC), the true Johnson and ST magnitudes ($V_0$ and $F555W_0$), the
{\sc allframe} photometric error on the ST mag ($\epsilon_{F555W}$),
the true Johnson-Cousins and ST colors ($(V-I)_0$ and $C_0^{\rm
HST}$), the photometric error on the ST color ($\epsilon_{C^{\rm
HST}}$), and the \noun{wfpc2} chip number (\#).  The complete
photometry is available in the electronic edition of the journal.

\subsection{Comparison with groundbased photometry }\label{sec:ground}

The zero-point of our $V$ photometry was checked against the
groundbased imaging of Hibbard et al. (\citeyear{hibbard_etal01}). 
There are only five stars which are not saturated in the \noun{wfpc2}
frames, and are bright enough to be measurable on the groundbased
frames. Total magnitudes were then found within an aperture of
$13\arcsec$, which was chosen after a growth curve analysis (the
seeing was $1\farcs 8$~FWHM). The average difference is $<V_{\rm
HST}-V_{\rm ground}> = 0.09 \pm 0.15$ for the five stars. Using only
the three stars with more regular growth curves, we find $<V_{\rm
HST}-V_{\rm ground}> =0.01 \pm 0.11$. The two magnitude scales are
then consistent within the mutual errors ($\zperrH01$~magnitudes for
the groundbased photometry, see Hibbard et
al. \citeyear{hibbard_etal01}).

We cannot make a similar test in $I$-band, since no suitable
groundbased photometry exists.  However, since the D00 recipe for the
calibration has no free parameters, we believe that the \noun{hst}
zero-point for the $I$ band is as reliable as that for the $V$ band.
Note that the same methods have given a good calibration of
\noun{wfpc2} $I$ photometry for each of $7$ globular clusters of the
Large Magellanic Cloud (Saviane et al. \citeyear{saviane_etal02}).

\subsection{NGC~625}

In order to constrain the distance to our field in the Antennae, we
compare our CMD to that of a well understood nearby galaxy with a
similar stellar population, for which deeper photometry and a well
measured CMD are available (see \S\ref{sec:distance}).  Toward this
end, we retrieved \noun{wfpc2} imaging of the dwarf irregular galaxy
NGC~625 from the \noun{hst} archive (Proposal ID=GO8708; PI=Skillman),
and $V,~I$ photometry was obtained following the methods described
above, and adopting $E_{B-V}=0.016$ (Schlegel et al. 
\citeyear{schlegel_etal98}).  The data consist of $4\times 1300$~s 
exposures in F555W and $8\times 1300$~s exposures in F814W, and thus 
reach limiting magnitudes comparable to those reached for NGC~4038/9.

\begin{figure*}

  \plotone{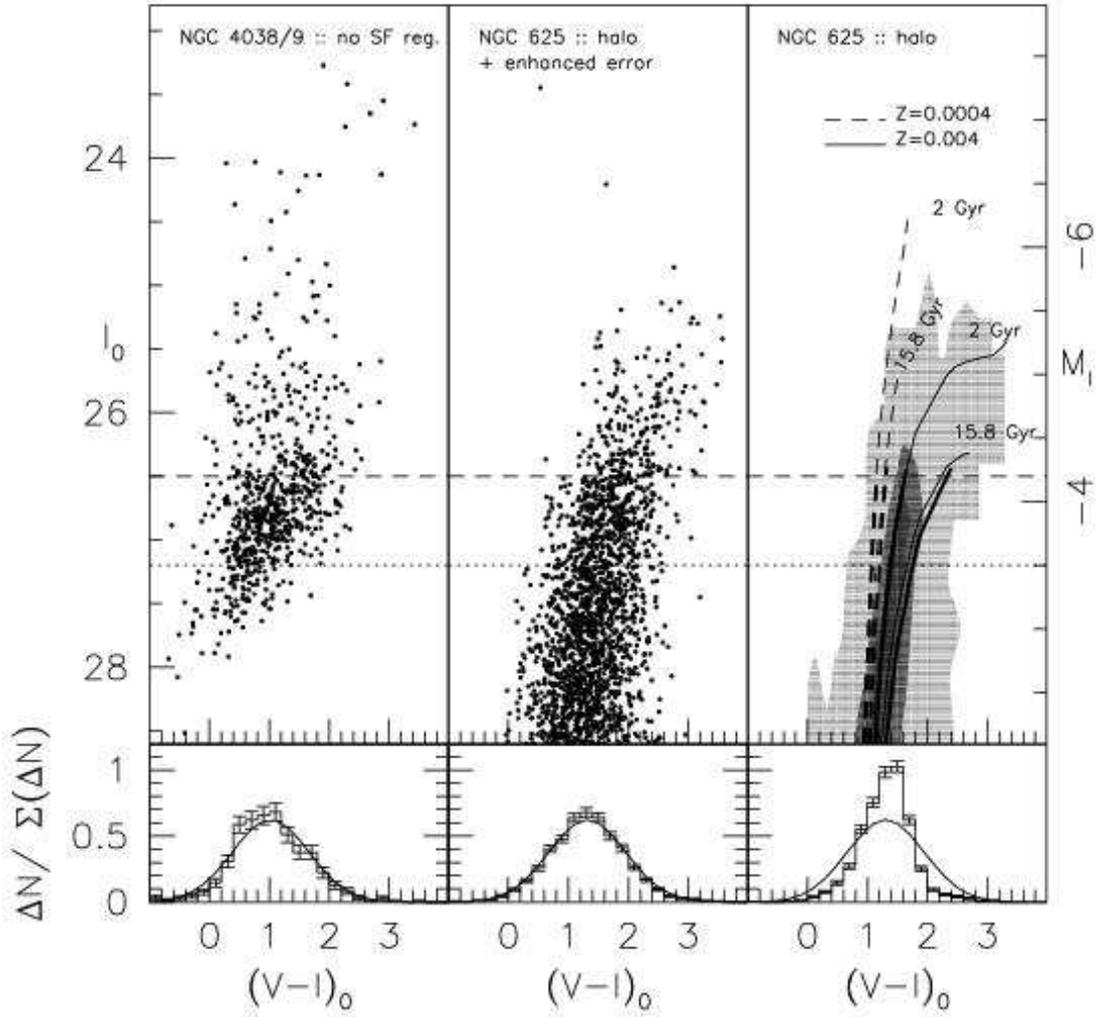}
\caption{The left and right panels show the CMDs of the NGC~4038 tail, and
NGC~625, after selecting stars far from the star formation regions and
adopting the distance moduli described in the text ($(m-M)_0=28.1$~mag
for NGC~625 and \mMo ~mag for NGC~4038/9). The central panel shows the 
CMD of NGC~625, after enhancing the photometric error. 
A set of isochrones from Girardi et al (\citeyear{girardi00}) is
overplotted to the NGC~625 diagram in the right panel ,
in order to illustrate the RGB (heavy lines) and AGB
evolutionary phases. A greyscale plot has been adopted for this CMD ($3$
levels, from $1$ to $130$ in steps of $43$ counts), in order to better
visualize the theoretical loci.
In each panel,
the dashed line 
marks the average luminosity of the RGB tip, $M_I=-4.2$.  The bottom  
panels show, for each CMD, the histogram of the color distribution, for 
stars brighter than $M_I=-3.5$ (dotted line). A Gaussian of 
$\sigma_{V-I}=0.64$ is plotted over each histogram. \label{fig:3cmd}}
\end{figure*}

\section{The distance to NGC~4038/9   } \label{sec:distance}

In order to estimate the distance to NGC~4038/9, we need to understand
the CMD of the NGC~4038 tail in terms of stellar evolutionary phases,
and thus to assign absolute luminosities to its stellar populations.
As we see in Fig.~\ref{fig:photerrors}, 
the
CMD is heavily affected by photometric errors, so its interpretation
is not straightforward. For this reason, we decided to compare it to
that of NGC~625.  This is a star forming dwarf irregular galaxy in the
Sculptor group, whose members have distance moduli in the range
$26$-$28$ (e.g. Jerjen, Freeman \& Binggeli \citeyear{jerjen_etal98}).
Thus, it should have a stellar population comparable to our region,
and we expect to see much better the various sequences, since its
distance modulus is $\sim 4$~mag smaller, and the exposure times are
almost the same.

\subsection{Interpreting the CMD of the NGC~4038 tail}
\label{sec:cmdsim}

Figure~\ref{fig:cal4} presents the color-magnitude diagrams constructed
for each of the four \noun{wfpc2} chips. The bright blue and red stars 
seen in the CMDs belong to the tail's youngest populations 
(see \S\ref{sec:space-age} below and Fig.~\ref{fig:pc_pops}).
There is also a rise in star counts below $I_0\sim26$, which could be
due either to the red giant branch (RGB) tip, or to the upper envelope
of the asymptotic giant branch (AGB), almost $1$~mag brighter.
The RGB of the oldest populations 
is typically the most conspicuous bright-red feature of the CMD of a
dwarf irregular galaxy (see e.g. Nikolaev \& Weinberg
\citeyear{nikolaev_weinberg00}). 
The AGB shares the color location of the RGB, extending to brighter
luminosities, but it is a shorter lived evolutionary phase and hence
less populated.
As is well known (e.g. Da Costa \citeyear{dacosta-canarie}), 
the luminosity of the RGB tip is almost independent of age and 
metallicity for ages greater than $2$~Gyr and for metallicities $0.0004
\leq Z \leq 0.004$. This produces a characteristic discontinuity in the
luminosity function which is used as a standard candle (see e.g. Da
Costa \citeyear{dacosta-canarie} and Carretta et
al. \citeyear{carretta_etal00}).
In the quoted metallicity range, we can assume a tip luminosity
$M_I^{\rm tip}=-4.2\pm 0.25$.

In order to understand whether the bright red stars of the NGC~4038
tail belong to the RGB or the AGB, we search for the best match
between the luminosity function (LF) of the old population of our
region, and that of the old population of NGC~625. However, we cannot
work with the original CMDs, because (1) the RGB/AGB of the old
population is polluted by the RGB of the young population, and (2) the
photometric errors in our CMD are much larger than those in NGC~625 at
comparable {\it absolute} luminosities. Thus, even if the underlying
theoretical LFs were the same, the RGB tip discontinuity for the
NGC~4038 tail would be less pronounced than that of NGC~625, and a
good match could never be made (see Madore \& Freedman \citeyear{bf95}
for a thorough discussion of the shape of the RGB LF vs. photometric
error and AGB contamination).

\begin{figure}

  \plotone{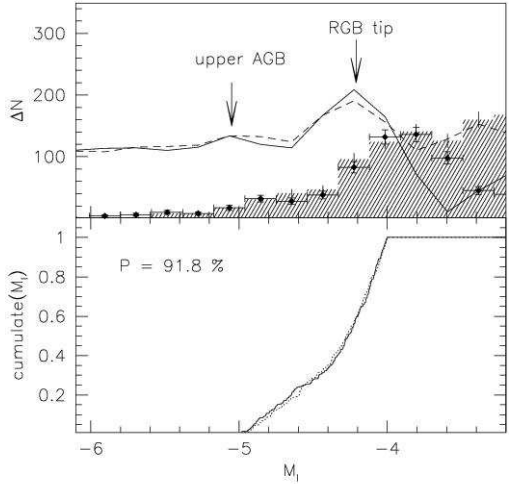}

\caption{Comparison of the LFs of NGC~4038 (filled diamonds and error 
bars) and NGC~625 (shaded histogram and error bars) for stars selected
far from star forming regions. 
The solid and dashed curves represent the results of applying a Sobel filter 
to the luminosity function of NGC~4048 and NGC~625, respectively.
The lower panel shows the cumulates of the NGC~4048 $M_I$ (solid line),
and that of the NGC~625 $M_I$, in the range $-5\leq M_I \leq -4$. The
result of the Kolmogorov-Smirnov test is also displayed.
\label{fig:match625}}
\end{figure}

In order to overcome the first difficulty, we attempt to isolate the 
CMD of the old population by selecting stars far from star formation
regions.  In the case of NGC~625, this means selecting the southern
$35\arcsec$ of the WF2 chip. For the NGC~4038 tail, we selected stars
in areas away from the star forming regions described in \S\ref{sec:intro}.  
The resulting isolated CMD for the NGC~4038 tail is given in the leftmost 
panel of Fig.~\ref{fig:3cmd}. 

To address the second difficulty, the errors of the NGC~625 photometry
were increased to match those of the NGC~4038 tail photometry, and then
the number of stars was reduced by a factor $r$ in order to match the
number of stars in the brightest part of the NGC~4038 tail CMD.  Thus we
added
random errors of
$\sigma_V=0.35$ to the $V$ photometry and $\sigma_I=0.25$ to the $I$
photometry of NGC~625, to match the errors on the NGC~4038 photometry at
$I_0\approx 26$.  The resulting degraded CMD for NGC~625 is given in the
middle panel of Fig.~\ref{fig:3cmd}. The lower panels of this figure
show histograms of the color distribution for stars brighter than
$M_I=-3.5$ (dotted line), in order to show that the photometrically
degraded NGC~625 data has a similar spread in color as the NGC~4038 tail
data.

Next, the two CMDs were compared for a range in the
difference of distance moduli, 
and for a range in $r$, until the best match was
found. The match was decided on the basis of a Kolmogorov-Smirnov (KS)
test on the two cumulated $M_I$ vectors, in the range $-5.0 \leq M_I
\leq -4.0$. 
For the comparison, the Whitmore et al. (\citeyear{whitmore}; hereafter
W99) distance modulus was adopted as a first guess for the distance of
the Antennae.
With this procedure, we derive a best fit of $r=2.4$ and 
a distance modulus difference of $2.6$~mag. More importantly, the best
fit is achieved when the discontinuity in the NGC~4038 tail LF coincides
with the RGB tip of the NGC~625 old population.
From the absolute magnitude of the RGB tip we then find a distance
modulus $(m-M)_0=\mMo \pm \errmMo$ for the Antennae, and in turn
$(m-M)_0=28.1\pm 0.25$ for NGC~625. 
The error is dominated by the uncertainty on the RGB tip luminosity
(which could be reduced if we knew the metallicity of the old
populations of the two galaxies). Instead, if we changed the difference
in the two distance moduli by only $\pm 0.08$~mag, the probability of
the coincidence as measured by the KS test would drop already below 
$90\%$.

Figure~\ref{fig:3cmd} shows the isolated CMD of NGC~625 shifted so as
to produce the best match to that of NGC~4038. 
The rightmost panel shows a greyscale
representation of the original CMD  of NGC~625 halo, plus some 
representative isochrones from Girardi et al. (\citeyear{girardi00}). 

The loci of the evolved RGB for 
two
ages and 
two
metallicities are drawn with heavy lines, while those for the AGB are
drawn with thin lines.
Furthermore, dashed and solid lines distinguish the two metallicities.
The isochrones illustrate several evolutionary features mentioned above.
Specifically, that the luminosity of the RGB tip is almost independent 
of age and metallicity for ages greater than $2$~Gyr and for 
metallicities $0.0004 \leq Z \leq 0.004$; that
the AGB shares the color location of the RGB, but has a brighter
termination and extends more to the red
for more metal rich populations; and most importantly that 
there is a sharp rise in star counts below $M_I\sim -4.2$, due to 
the more densely populated RGB.

\begin{figure}

  \plotone{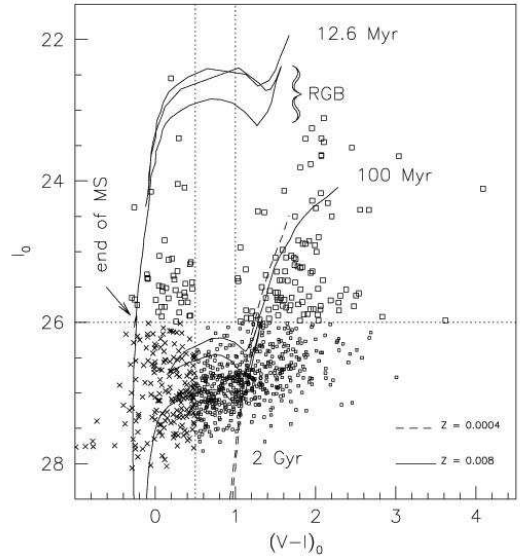}

\caption{The CMD of the PC field with three representative
isochrones, assuming $(m-M)_0=\mMo$. This diagram is used to define four
subpopulations of stars: bright and faint blue MS stars younger and 
older than $\sim 12$~Myr (crosses and the large squares in the upper 
left region), bright red RGB and core helium burning stars younger 
than $\sim 100$~Myr (large squares in the upper right region), and 
faint red RGB and AGB stars older than $\sim 2$~Gyr (small squares).
The dotted lines separate these four populations. 
Note that the vertical extent of the bright red region is larger than
that of a single RGB (identified by the curly bracket), a sign of
extended star formation.
\label{fig:pc_pops}}
\end{figure}

The close match between the two CMDs is confirmed in
Figure~\ref{fig:match625}, where we display the two LFs, and the result
obtained with the convolution with a Sobel edge detector (see Lee,
Freedman \& Madore \citeyear{lfm93}). We see the coincidence of both
the main and secondary peaks of the two curves, where the higher
maximum corresponds to the RGB tip, and the lower maximum corresponds
to the upper envelope of the AGB. The lower panel of the figure
confirms the coincidence in a quantitative way: the probability that
the the two distributions in $M_I$ share a common parent is $91.8~\%$.
Based on this experiment, we thus conclude that NGC~4038/9 is at a
distance of $\DMpc \pm \errDMpc$~Mpc.

\begin{figure*}
  \plotone{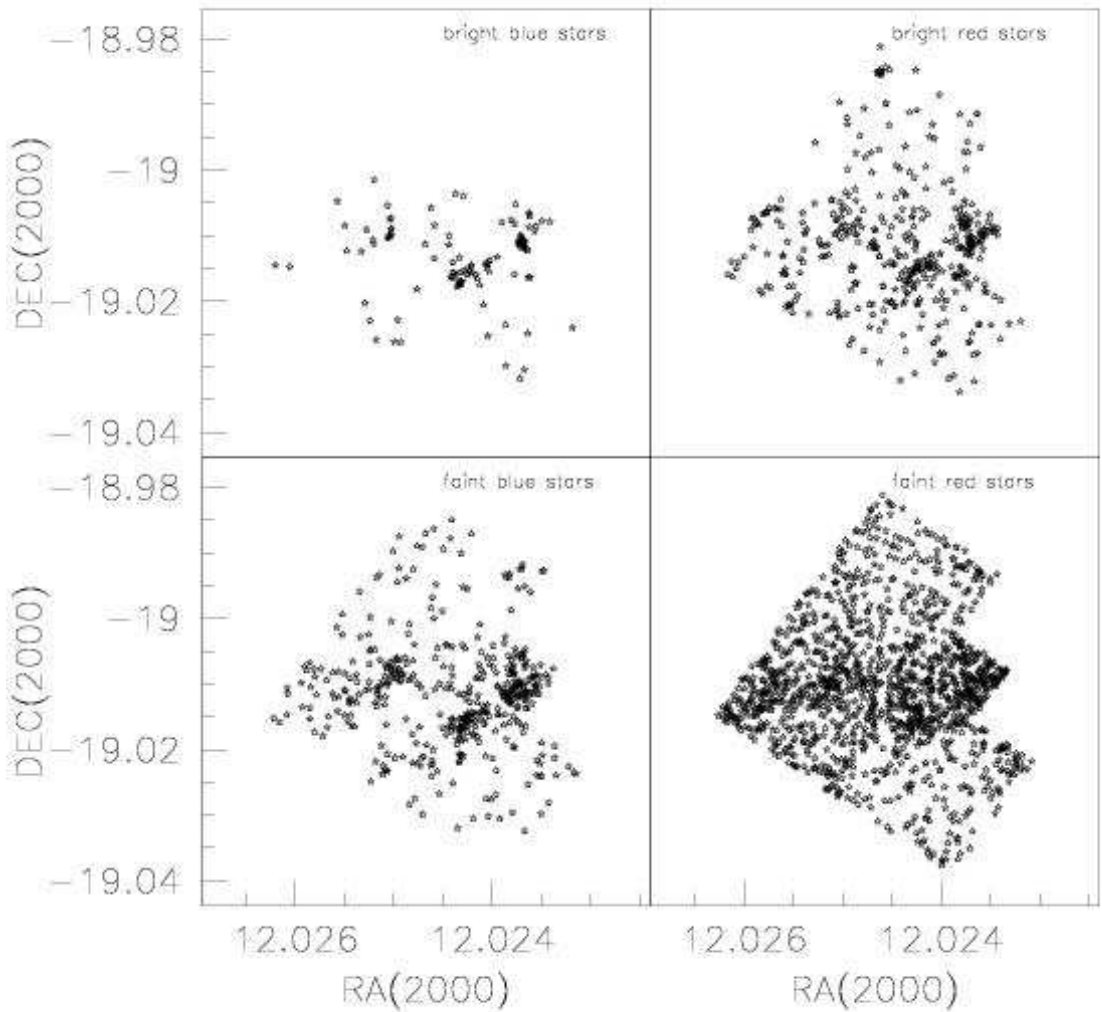}
\caption{J2000 positions of several subpopulations, identified by the
labels in the upper right corner. These populations are those
delineated in Fig.~\ref{fig:pc_pops}. The faint red stars, which 
we identify with old stars originally in the outer disk of NGC~4038, 
show a broader and uniform spatial distribution. 
\label{fig:radec}}
\end{figure*}

The Antennae is not the most distant system for which a distance based
on the RGB tip has been attempted: Harris et al. (\citeyear{vcc1104})
used this method to find the distance of a dwarf elliptical galaxy of
the Virgo cluster (at $\sim 16$~Mpc).  In order to obtain the deepest
possible LF, Harris et al. imaged the galaxy with the HST only in the
$I$ filter. Thus, it is not possible to carry out a detailed
comparison of the CMDs, as we did for NGC~625. On the other hand,
the photometry of Harris et al. has  photometric errors comparable to 
ours, permitting us to compare the two without any photometric 
manipulation.  In appendix \ref{sec:vcc1104} we show that the close 
similarity of the two LFs adds further support to our distance modulus.

\subsection{Comparison with previous distance determinations} 
\label{sec:otherdistances}

All recent studies of the Antennae adopt a distance based on its
redshift, but the results are not unique, even if the adopted
heliocentric velocity is almost the same ($v_{\rm
hel}\approx1650$~km~s$^{-1}$): variations of almost a factor of two
are found in the literature (from e.g. $19$~Mpc, in Wilson et
al. \citeyear{wilson00}, to $33$~Mpc in MDL92).  In fact, different
authors can adopt different values for the Hubble constant, and they
can decide to correct for the solar motion in different ways
(i.e. with respect to the Galactic center, to the Local Group, or no
correction at all).

\begin{figure*}
  \plotone{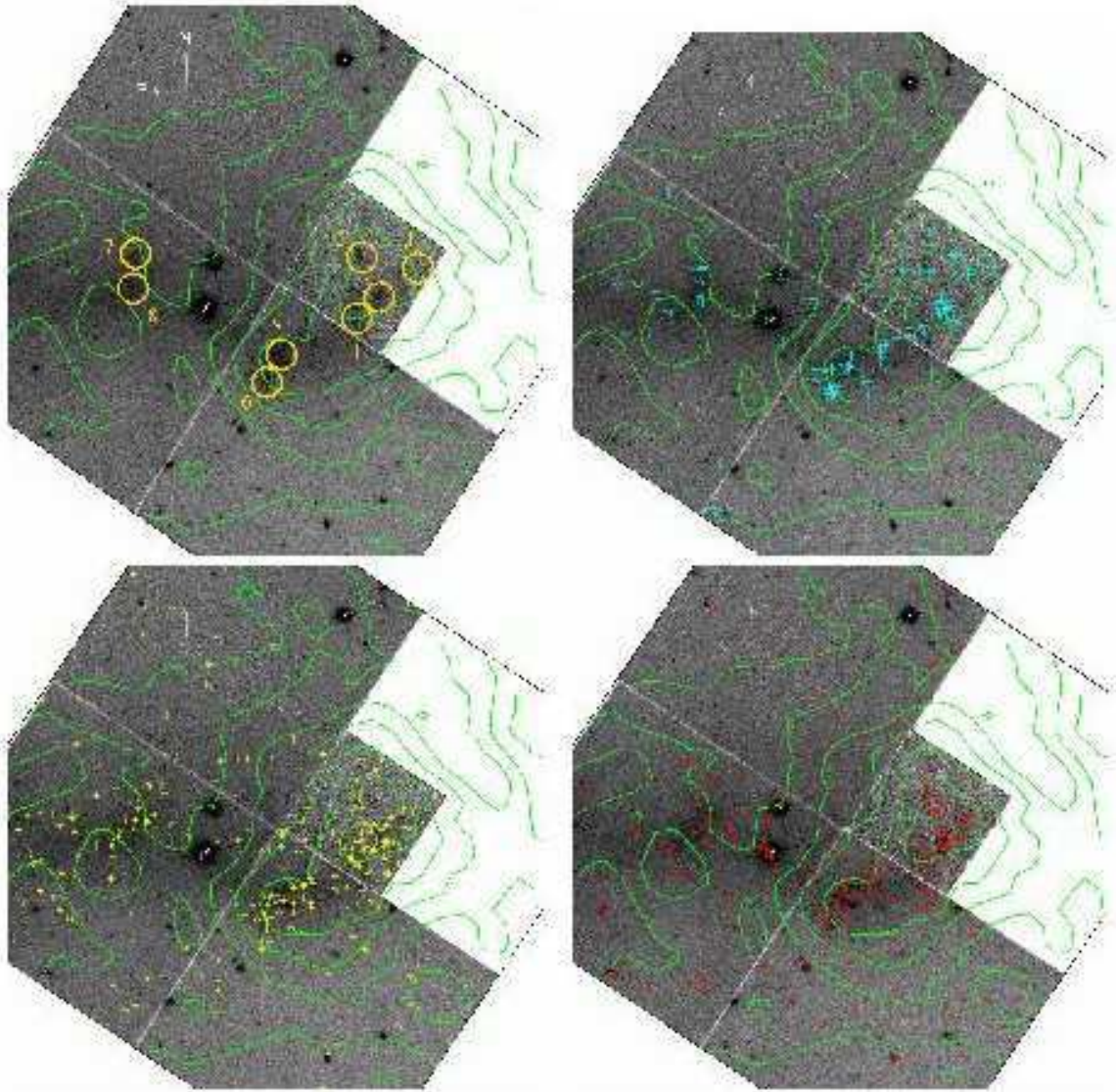}

\caption{
Greyscale representation of the F555W
frame with H{\sc i} column densities from the intermediate resolution 
datacube of Hibbard et al. (2001) and the location of various stellar 
populations (see Fig.~\ref{fig:pc_pops}) superimposed. The H{\sc i} 
density runs from  $5\times10^{19}~\rm cm^{-2}$ to \protect\( 
1.05\times 10^{21}\, \rm cm^{-2}\protect \), in steps of $2\times 
10^{20}~\rm cm^{-2}$.  
{\em Top left panel}: The location of the eight blue stellar 
associations identified in Table~\ref{tab:associations} and discussed 
in \S\ref{sec:recent-sf} are indicated by the numbered circles with 
a radius of $4''$. {\em Top right panel}: As in the first panel, but 
with the location of the bright blue stars indicated by cyan crosses. 
{\em Bottom Left panel}: As in the first panel, but 
with the location of the faint blue stars 
indicated by yellow crosses. 
{\em Bottom Right panel}: As in the first panel, but 
with the location of the bright red stars 
indicated by red crosses. 
\label{fig:hst-hi}}
\end{figure*}

Therefore, a distance based on a standard candle is in dire need for
this object, which is considered to be of great interest by a diverse
community.  In the past, this was attempted by Rubin, Ford \&
D'Odorico (\citeyear{rubin}), who were among the first to measure the
radial velocity of the system ($v_{\rm hel}=1636$~km~s$^{-1}$).  They
corrected the velocity to a Galactocentric value $v_{\rm
rad}=1464$~km~s$^{-1}$, and computed a distance of $19.5$~Mpc (for
$H_0=75$~km~s$^{-1}$~Mpc$^{-1}$). However, they disregarded it, since
``for such a low value of the redshift, the effects of anisotropy in
the velocity field make the determination of distance by means of the
Hubble constant unreliable''.  Instead, they used as their primary
distance indicator the size of the disk H\noun{ii} regions, which gave
$D=6$~Mpc. Other distance indicators (the brightest stars and the 1921
supernova) gave distances ranging from $6$~Mpc to $13$~Mpc, due to the
unknown amount of internal absorption. The authors also pointed out
that NGC~4027 is likely physically linked to the pair, and the
distance of the latter galaxy at the time was estimated
\( 12.5\pm 2.5 \) Mpc (de Vaucouleurs et al. \citeyear{ngc-4027}). 
In conclusion, although an accurate determination was not possible,
there were indications of a distance shorter than that predicted by
the redshift.

If, based on the Rubin et al. argument, we don't put much confidence
in the redshift distance, then our new distance determination is
within the range of values discussed above. This means that the
velocity of the galaxy is enhanced with respect to a pure Hubble flow,
and we can see if this large peculiar velocity is compatible with
modern flow models.

A recent, detailed model predicting the velocities of galaxies in the
nearby Universe can be found in Tonry et
al. (\citeyear{tonry_etal00}), who also provide the software to make
one's own
calculations\footnote{http://www.ifa.hawaii.edu/\textasciitilde{}jt/SBF/sbf2flow.f}.
The input of the model is the Supergalactic position of the object.
If we assume our short distance, then the position of the Antennae is
$(x,y,z)_{\rm SGB}=(-8.78, 9.83, -4.09)~\rm Mpc$. At that position,
the model predicts a radial velocity $v_{\rm rad}^{\rm mod}=1496 \kms$
in the cosmic microwave background (CMB) reference frame (and actually
the total predicted velocity is almost aligned with the line of
sight). The velocity of the Antennae in the same reference frame is
obtained by adding the velocity of the Sun with respect to the CMB,
which is $369 \kms$ in the direction $(SGL, SGB)=(116.72, -26.69)$
(Lineweaver et al. \citeyear{lineweaver_etal96}). Its projection along
the line of sight is $353.21 \kms$, so finally we obtain a velocity of
the Antennae in the CMB reference frame of $v_{\rm rad}^{\rm
obs}=1995.21 \kms$. The observed velocity is thus $500 \kms$ larger
than the predicted one. The Tonry et al. model
predicts a peculiar velocity dispersion of $187 \kms$ with respect to
the local flow, so at our shorter distance the velocity of NGC 4038/9
is $2.7~\sigma$ larger than predicted by the model. While this is
not an extreme discrepancy, one would still like to understand its
origin. The discrepancy would be $+330\kms$ at the W99 position, and
$-660\kms$ at the MDL92 position.

\begin{figure}

  \plotone{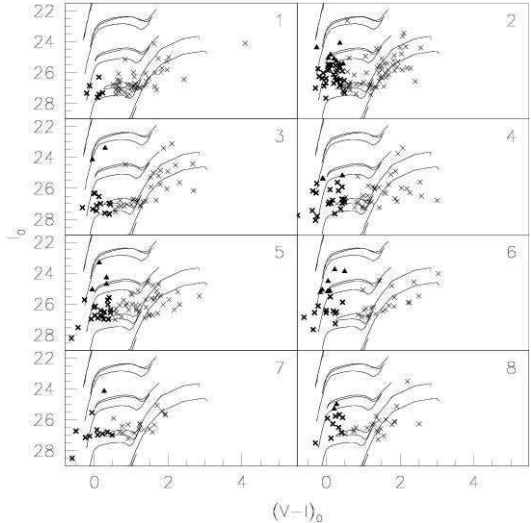}

\caption{Color-magnitude diagrams of \protect\( 8\protect \) 
associations of blue stars.  The solid curves represent post-MS
isochrones taken from Girardi et al. (\citeyear{girardi00}), for ages
of \protect\( 4\protect \), \protect\( 12.6\protect \), \protect\(
39.8\protect \), \protect\( 126\protect \), and \protect\( 398\protect
\) Myr (corresponding to masses at the termination of the MS of about 
\protect\( 42.6\protect \), \protect\( 15.8\protect \), \protect\( 
8\protect \), \protect\( 4.5\protect \), and \protect\( 2.7\, M_{\odot
}\protect \)), and a metallicity ca. \protect\( 1/2\protect \) solar
(\protect\( Z=0.008\protect \)).  Here we adopt \protect\(
(m-M)_{0}=\mMo\protect \). The groups of stars have been selected from
within a radius of \protect\( 4\arcsec \protect \) around sites of more
active star formation.  The plots are arranged in order of increasing
RA, from left to right and top to bottom. Thus the top \protect\(
4\protect \) panels show CMDs from the PC chip star formation
regions. They are followed by \protect\( 2\protect \) regions taken from
the WF4 chip, and finally the bottom \protect\( 2\protect \) come from
the WF3 chip. The stars bluer than \protect\( (V-I)_{0}=0.5\protect \)
(heavier crosses) have been used to construct the LFs, 
and among these, potential
ionizing stars 
(i.e. $M_I\leq-5.23$) 
are plotted as filled 
triangles. \label{fig:assoc-cmds}}
\end{figure}

We can recall that the short distance found by Rubin et al. was in
agreement with simple flow models of the time. In particular, using the
kinematic model of the local supercluster of galaxies by de Vaucouleurs
(\citeyear{supercluster58}), Rubin et al. predicted a distance $D=(1\pm
0.1)\times D_{\rm V}=10\pm 1$~Mpc, where $D_{\rm V}$ is the distance of
the Virgo cluster.  In the de Vaucouleurs (\citeyear{supercluster58})
model, the local supercluster had both a differential rotation and a
recession velocity from its center, located in Virgo.  The high velocity
of the Antennae was then explained by its rotational velocity around the
Virgo cluster center.  Note that the distance to the Virgo cluster is
now estimated \( D_{\rm V}=16.4 \)~Mpc (Ferrarese et al.
\citeyear{laura-m87}), so the model would now give a distance to
the Antennae of \( D\simeq 15 - 18 \) Mpc, and a revision of the model
parameters 
might bring $D$ into agreement with our smaller value. 

Another aspect of the problem is that flow models are valid in a
statistical sense, and for relatively large portions of the Universe,
so they can break down for individual objects.  Indeed, the same
discrepancy seen for the Antennae is present for one of the Tonry et
al. galaxies near the position of the Antennae. Looking at their
Fig.~11, one notes an object near $(x, y)_{\rm SGB}=-10,10~\rm Mpc$
whose radial velocity is, again, $\sim 300 \kms$ larger than that
predicted by the model. Objects near the Antennae region start to feel
the pull of the Great Attractor (GA), toward $SGL\sim 160$~deg, and
Fig.~11 of Tonry et al. shows several other objects with residual
velocities pointing toward the GA.

A final indication of a short distance is given by the distance to
NGC~4027, if we still accept the physical association of the two
objects.  The value is still uncertain, but Pence \& de Vaucouleurs
(\citeyear{dV-4027}) adopt \( 10\pm 1.5 \) Mpc from several distance
indicators. Again, this value is smaller than that based on the
recession velocity (\(D\approx 14\)~Mpc).

\begin{figure}

  \plotone{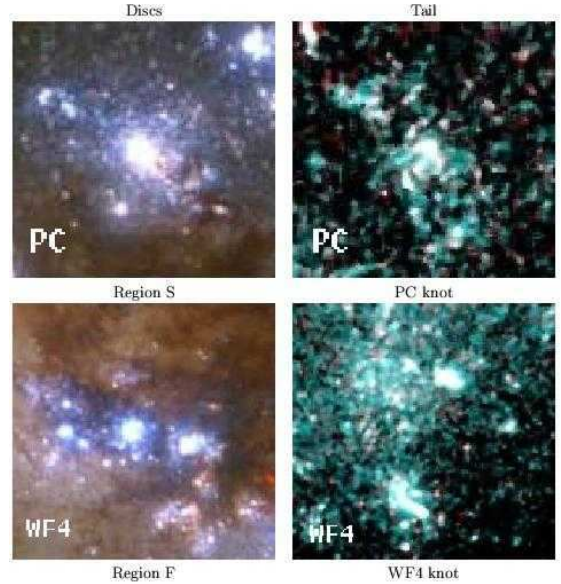}

\caption{
Here we compare the visual appearance of two OB associations of the
tidal tail to that of two `super' clusters extracted from Whitmore et
al. (\citeyear{whitmore}). The size of the fields are $\wfside \times 
\wfside$ and $\pcside \times \pcside$ arcsec for the WF4 and PC images, 
respectively.
\label{fig:associations}}
\end{figure}

We note that the Tully ``Nearby Galaxies Catalog'' (Tully
\citeyear{TullyCatalog}) places NGC~4038/9 in group 22-1 at a
distance of $21$~Mpc (based on distances to $5$ members of the group).
The same value has been found for NGC~4033 in that group by Tonry et
al.  (\citeyear{tonry4033}).  There are regions of the sky where
structures with similar redshifts but different distances can overlap,
but this is not the case for the Antennae, which are closer than the
Great Attractor and away from the Virgo cluster (look at Fig.~20 of
Tonry et al., at position $(x, y)_{\rm SGB}\approx (-10,10)$~Mpc).  If
our distance is confirmed, we must conclude that the association of
the Antennae with group 22-1 is an optical superposition, as the
Antennae are a factor $1.5$ closer.

\section{The stellar content of the southern tidal tail  }
\label{sec:star-content}

Insight into the star formation history of the tidal tail can be
obtained from both the spatial distribution of stars selected from
different regions of the CMD, and from direct analysis of the CMD. In
the following sections we first analyze the spatial distribution of
several subpopulations of stars, and then perform a more detailed
comparison of the CMD of the young population to the theoretical
isochrones.  We then estimate the mass in young stars by several
methods, using the main sequence luminosity functions of the star
forming associations. Finally, we estimate the mass in old stars by
comparing the luminosity function of stars located far from star
forming regions to that of a dwarf spheroidal galaxy of the Local
Group.

\begin{figure*}

  \plotone{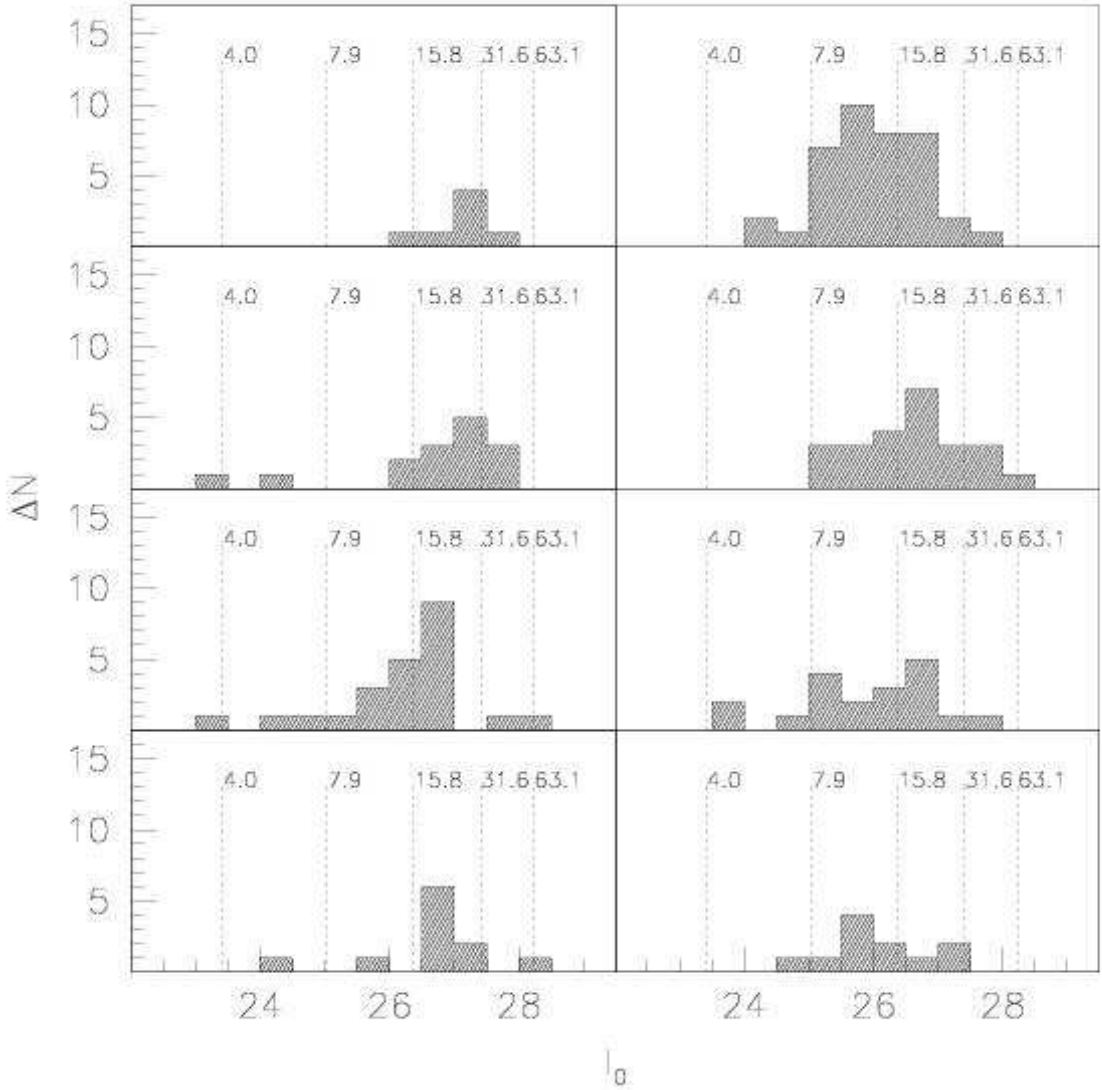}

\caption{The luminosity function of the blue stars identified 
in Fig.~\ref{fig:assoc-cmds}. The vertical dotted lines mark the 
end of the MS phase for several choices of the age, in Myr, and a 
metallicity \protect\( Z=0.008\protect \). The arrangement of the 
panels is the same as that of the CMDs. \label{fig:assoc-lfs}}
\end{figure*}

\subsection{Spatial distribution of populations as a function of age 
}\label{sec:space-age}

Fig.~\ref{fig:pc_pops} shows the CMD 
from
the PC chip, and three
isochrones from Girardi et al. (\citeyear{girardi00}) which have been
placed at the distance of the Antennae for $(m-M)_0=\mMo$. The
$Z=0.0004$ isochrone (dashed line) represents the metal-poor,
intermediate and old population. The two $Z=0.008$ isochrones show the
CMD morphology of the younger populations.   For these isochrones,
we use the nebular oxygen abundance of  $\approx 1/3$~solar measured
by MDL92 which likely applies to all of the recently formed stars.
Older stars likely have lower metallicities, so in the following 
we let $Z$ vary within a wide range (from $0.004$ to $0.02$).

Using Fig.~\ref{fig:pc_pops}, we define four regions of the CMD:
bright blue, faint blue, bright red and faint red stars. The limit
between bright and faint is set at $I_0=26$. Blue stars are stars
bluer than $(V-I)_0=0.5$. Faint red stars are those redder than
$(V-I)_0=0.5$, while bright red stars are those redder than
$(V-I)_0=1$. These regions are delineated by dotted lines in
Fig.~\ref{fig:pc_pops}.  For the quoted distance modulus, and letting
the metallicity $Z$ vary between $Z=0.0004$ and $Z=0.02$, 
we find that the regions delineating the bright and faint blue stars
will contain mostly MS stars of ages less than or greater than
$\approx 10$ to $\approx 13$~Myr, depending on
$Z$. In Fig.~\ref{fig:pc_pops} we see that the ``hook'' marking the
end of the MS is right at $I_0=26$ for a $12.6$~Myr, $Z=0.008$
population. The region of bright red stars will contain mostly RGB and
core helium burning stars younger than $\approx 80$ to $\approx
100$~Myr, again depending on $Z$. The limit $I_0=26$ is reached by
the RGB tip of a $100$~Myr, $Z=0.008$ isochrone; the isochrone also
shows that the progenitors of bright red stars were located inside the
faint blue region. Finally, the faint red stars region will contain
mostly RGB and AGB stars older than $\approx 2$~Gyr.

\begin{deluxetable*}{lccccccccccc}
\tablecolumns{13} 
\tablewidth{0pc} 
\tablecaption{Photometric properties of the \protect\( 8\protect \) 
associations. \label{tab:associations}  }
\tablehead{ 
\colhead{} &
\colhead{} &
\colhead{} &
\colhead{} &
\colhead{} &
\multicolumn{3}{c}{{blue }}&
\colhead{} &
\multicolumn{3}{c}{{total }}\\
\cline{6-8} \cline{10-12}
\colhead{\#} &
\colhead{\( RA \)}&
\colhead{\( DEC \)}&
\colhead{\( N_{\rm b} \)}&
\colhead{\( N_{\rm T} \)}&
\colhead{\scriptsize \( M_{I} \)}&
\colhead{\scriptsize \( M_{V} \)}&
\colhead{\scriptsize \( (V-I)_{0} \)}&
\colhead{} &
\colhead{\scriptsize \( M_{I} \)}&
\colhead{\scriptsize \( M_{V} \)}&
\colhead{\scriptsize  \( (V-I)_{0} \)}\\
}
\startdata 
1&
{ 12:01:24.53 }&
 {-19:00:32.40 }&
 {7 }&
 {54 }&
 {-6.4 }&
 {-6.6 }&
 {-0.1 }&
 {  } &
 {-9.6 }&
 {-8.3 }&
 {1.2 }\\
2$\equiv$III&
 {12:01:25.30 }& 
 {-19:00:40.32 }&
 {39 }&
 {96 }&
 {-9.7 }&
 {-9.5 }&
 {0.1 }&
 {  } &
 {-10.9 }&
 {-10.1 }&
 {0.8 }\\
3&
 {12:01:25.63 }& 
 {-19:00:28.80 }&
 {15 }&
 {47 }&
 {-8.8 }&
 {-8.6 }&
 {0.1 }&
 {  } &
 {-10.2 }&
 {-9.2 }&
 {1.0 }\\
4&
 {12:01:25.78 }& 
 {-19:00:45.37 }&
 {24 }&
 {60 }&
 {-8.5 }&
 {-8.5 }&
 {0.1 }&
 {  } &
 {-9.9 }&
 {-9.0 }&
 {0.9 }\\
5$\equiv$I&
 {12:01:27.29 }& 
 {-19:00:54.72 }&
 {23 }&
 {64 }&
 {-9.3 }&
 {-9.1 }&
 {0.2 }&
 {  } &
 {-10.2 }&
 {-9.5 }&
 {0.7 }\\
6$\equiv$II&
 {12:01:27.57 }& 
 {-19:01:01.56 }&
 {19 }&
 {41 }&
 {-9.3 }&
 {-9.2 }&
 {0.1 }&
 {  } &
 {-10.0 }&
 {-9.4 }&
 {0.6 }\\
7&
 {12:01:30.13 }& 
 {-19:00:26.28 }&
 {11 }&
 {29 }&
 {-8.0 }&
 {-7.9 }&
 {0.1 }&
 {  } &
 {-9.0 }&
 {-8.3 }&
 {0.7 }\\
8&
 {12:01:30.19 }& 
 {-19:00:35.28 }&
 {11 }&
 {32 }&
 {-8.1 }&
 {-7.8 }&
 {0.2 }&
 {  } &
 {-9.3 }&
 {-8.3 }&
 {1.0  }\\
\enddata 
\tablecomments{
The absolute magnitudes have been computed assuming 
the W99 distance modulus (\protect\( m-M_{0}=31.4\protect \)).  
Columns \protect\( RA\protect \) and \protect\( DEC\protect \) 
are the equatorial J2000 coordinates, \protect\( N_{\rm b}\protect \) 
is the number of blue stars, and \protect\( N_{\rm T}\protect \) is 
the total number of stars. 
Magnitudes and colors have been corrected only for foreground absorption
and reddening.
The associations number is given in the
first column, and the labels I, II, and III identify the three star 
forming regions studied by MDL92.}
\end{deluxetable*}

The spatial distribution of each of these populations is shown in
Figs.~\ref{fig:radec} and \ref{fig:hst-hi}. Fig.~\ref{fig:hst-hi} also
shows the distribution of the starlight (greyscales) and the neutral
atomic gas (contours; from the VLA\footnote{The Very Large Array (VLA)
of the National Radio Astronomy Observatory is operated by Associated
Universities, Inc., under cooperative agreement with the National
Science Foundation.} data of Hibbard et al. \citeyear{hibbard_etal01}).
With respect to the gas distribution, we see that the stars are well
embedded within the ISM, with no signs of a relative displacement. 
Interestingly, a higher H{\sc i} density is observed in coincidence 
with the eastern bar-like structure, where the most active
star formation is observed (see \S\ref{sec:recent-sf}), and which
contains the three H{\sc ii} regions studied by MDL92. The youngest
(bright blue) stars are clearly also concentrated along this
structure. Moreover, the most active sites of star formation are
located at the two extremes of the elongated structure, as is often
observed in dwarf irregulars. Not surprisingly, both luminous and
fainter main sequence stars follow a similar spatial distribution,
with the youngest stars being the most tightly concentrated.

Bright red stars follow the distribution of main sequence stars, which
indicates that the star formation started at least a hundred Myr ago.
The faint red stars, which we identify with old stars originally in
the outer disk of NGC~4038, show a broader and more uniform spatial
distribution (Fig.~\ref{fig:radec}), which gives an idea of the
underlying structure of the tail. Notice also an apparent increase in
surface density toward the North, probably connected with the low
surface brightness extension of the tail first discussed by S78 (see
also Fig.~\ref{fig:dss}).

\subsection{Young star forming associations}\label{sec:recent-sf}

In order to characterize the recent star formation, we have selected
those stellar associations which show an overdensity of stars within a
few arcseconds, and which contain stars brighter than
$I_0=27.5$~magnitudes (i.e. younger than $32$~Myr for $Z=0.008$ and our
choice of distance modulus).  Magnitudes and star counts were measured
within a radius of $4\arcsec$, which corresponds to $\simeq 268$~pc for
a distance modulus $(m-M)_0=$ \mMo, or $\simeq 370$~pc for the more
commonly accepted distance modulus of $(m-M)_0=31.41$ (e.g. W99). The
result is \(8 \) associations, whose locations are indicated in the top
left panel of Fig.~\ref{fig:hst-hi}, and whose CMDs are shown in
Fig.~\ref{fig:assoc-cmds}.

The CMDs of these \(8 \) associations are used to compute their
integrated properties, which are summarized in
Table~\ref{tab:associations}. The table presents, from left to right,
our identification number (and that of MDL92 when appropriate), the
equatorial coordinates of the center of each region, the number of blue
and total number of stars, then separately for the blue \( (V-I)_{0}\leq
0.5 \) and total populations: the absolute integrated \( I \) and \( V
\) magnitudes and true color.  Not surprisingly, the three associations
with the highest flux in $V$ (i.e. those having $M_V\leq-9.4$) coincide
with the three H{\sc ii} regions studied by MDL92.

We now address the question of whether these regions contain enough
ionizing stars to account for the H$\alpha$ fluxes observed by MDL92. To
produce a significant ionizing flux, stars must be earlier than O5V,
i.e. brighter than $M_I=-5.23$ (Cox \citeyear{allen}). We therefore
select stars bluer than 
\( (V-I)_{0} = 0.5 \)
and brighter than
$I_0=25.47$ 
in our CMD (drawn with black triangles in Fig.~\ref{fig:assoc-cmds}).
We estimate the expected Lyman continuum luminosity ($L_{\rm Lyc}$) of
these stars from Table~3 of Vacca et al. (\citeyear{vacca_etal96}). Given
the Lyman continuum luminosities of each region, we calculate the
expected H$\alpha$ luminosities using Table~2 of
Rozas et al. (\citeyear{rozas_etal99}). The expected H$\alpha$ luminosities are
$9\times 10^{38}$~erg~sec$^{-1}$, $1.1\times 10^{39}$~erg~sec$^{-1}$,
and $6.8\times 10^{38}$~erg~sec$^{-1}$ for regions I, II, and III,
respectively. Comparing these to the H$\alpha$ luminosities measured by
MDL92, converted to our smaller distance ($5\times
10^{38}$~erg~sec$^{-1}$, $1.4\times 10^{38}$~erg~sec$^{-1}$, and
$2.8\times 10^{37}$~erg~sec$^{-1}$ for regions I, II, and III
respectively), we find that that there is more than enough ionizing
radiation expected in these regions.

Another clue to the nature of the recent star formation is provided by
a comparison of our modest sample of clusters with the so-called Super
Star Clusters (SSCs) observed throughout central regions of the
merging disks by Whitmore et al. (\citeyear{whitmore}). Figure
\ref{fig:associations} shows such a side-by-side comparison, where the
left column includes two examples of the W99 superclusters, and the
right column includes two instances taken from our frames, where the
three brightest knots are displayed (Regions I-III of MDL92).  In
order to compare the star formation regions at different resolutions,
the top two images have been extracted from the PC chip ($0\farcs05$
pixels), whereas the bottom two come from the WF4 chip ($0\farcs10$
pixels).

Within the central regions, there are dozens of roughly spherical SSCs
which contain more than $10^5$~M$_\odot$ of stars (W99). In contrast,
the associations observed in the tidal tail are irregular in shape and
contain less than a hundred stars brighter than $I_0=28$. Indeed, we
see from Table~\ref{tab:associations} that even the brightest
associations have \( V \) luminosities falling in the faint bin of the
W99 cluster LF, and most are close to their \( M_{V}=-9 \) faintest
limit.  It is then clear that within the tail we do not observe the
kind of SSCs that have been produced in the ``violent{}'' central
regions. The linear sizes look comparable to those of the central
clusters, implying that the tail associations are in a much less dense
environment. One might conclude that SSCs require the higher pressures
found in the central regions in order to form (e.g., Jog \& Das
\citeyear{jogDas92}, \citeyear{jogDas96}; Elmegreen \& Efremov 
\citeyear{elmegreen97}), while spontaneous star formation in the 
tail produces the kind of O-B star associations seen in dwarf 
irregular galaxies.

\subsection{Comparison to theoretical isochrones} \label{sec:comp-isos}

The CMDs for each of the associations are presented in
Fig.~\ref{fig:assoc-cmds}, with post-MS isochrones taken from Girardi et
al. (\citeyear{girardi00}) drawn for ages between $4$ and $400$ Myr (for
a distance modulus of $\mMo$), and a metallicity $Z=0.008$.
Figure~\ref{fig:assoc-lfs} shows the luminosity functions for each
association, with the luminosity of the termination of the MS (TMS)
indicated by vertical dashed lines, for a range of ages and the same
metallicity.  Note that adopting the larger W99 distance modulus would
shift all ages to younger values. For example, most stars of association
$\#2$ would be older than $6$~Myr, instead of $8$~Myr.

The CMDs show a distinct blue MS 
up to
$\sim 40$~Myr, but also a
population of red supergiants, which are stars in either the RGB or
the core helium burning phase. Since the bright red stars follow the
spatial distribution of the blue stars (Figs.~\ref{fig:radec} and
\ref{fig:hst-hi}), we believe that the red populations belong to the
same associations defined by the blue stars.  If these associations
represented single star formation episodes, then the RGB stars would
be confined in an almost vertical CMD region with a relatively short
extent in luminosity ($\sim 1$~mag) 
as
shown in
Fig.~\ref{fig:pc_pops}. Instead, the red vertical sequence extends for
several magnitudes, as one can see in the most populated associations
like $\#2$. Thus, it seems like stars have been forming for a few
hundred Myr, possibly at a continuous rate.

The most populous association is $\#2$ (Region III from MDL92, also
shown in the upper right panel of Fig.~\ref{fig:associations}), with
\( 39 \) stars on the MS, and its CMD shows that most of its stars
were formed more than \( \sim 8 \) Myr ago (see also the upper right
panel of Fig.~\ref{fig:assoc-lfs}). Both the CMDs and LFs seem to
indicate that there are few stars younger than $8$~Myr in any of the
eight associations.  However, it is interesting to note that the
youngest stars are observed in associations $\#2$, $3$, $5$, and $6$,
and that three of these coincide with the star formation regions
studied by MDL92. In particular, the youngest stars are seen in
association $5$, which coincides with association I of MDL92, who
indeed found that this association is younger than the other two,
containing stars whose age they estimated as $\leq 2$~Myr. We find
stars as young as $3$~Myr in that association, and 
using an isochrone with slightly lower metallicity would reconcile our
estimate with that of MDL92. For the other two associations, MDL92
estimated ages $\leq 6$~Myr, and indeed stars younger than that age are
observed in associations $2$ and $6$.

\begin{figure}

  \plotone{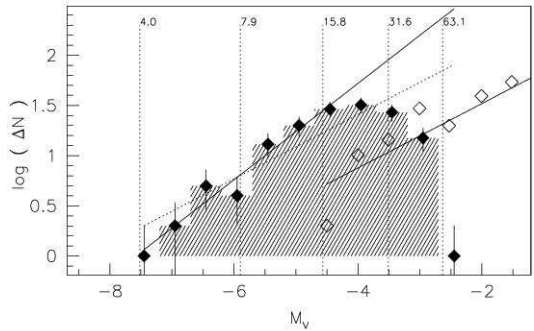}

\caption{Here we show the 
logarithmic integrated \protect\( M_V\protect \) luminosity function
of all the \protect\( 8\protect \) associations (filled diamonds and
shaded histogram). The solid line is a power law with exponent
\protect\( 0.48\protect \), as obtained by Cole et
al. (\citeyear{cole_etal99}) for IC~1613, fitted to the data for the
brightest two magnitudes (\protect\( V_{0}\leq 26.5\protect \) or
$M_V\leq -4.2$). The open diamonds represent the main sequence LF of the
LMC cluster NGC~2004 (Keller et al. \citeyear{keller_etal00}), to
which a straight line has been fitted, yielding a slope
\( a=0.32\pm 0.04 \).  The dotted line represents a power law of the 
same exponent, fitted to the bright part of our LF.  We assumed a LMC 
distance modulus and absorption of \protect\( (m-M)_{0}=18.5\protect \) 
and \protect\( A_{V}=0.249\protect \) respectively. The vertical dotted 
lines mark the end of the MS phase for several choices of the age, in 
Myr, and a metallicity \protect\( Z=0.008\protect \). With these choices, 
we find the same age for NGC~2004 as that derived by Keller et al
(\citeyear{keller_etal00}), i.e. \protect\( 15.8\protect \) Myr.
\label{fig:lf-8-log}
}

\end{figure}

\begin{figure}

  \plotone{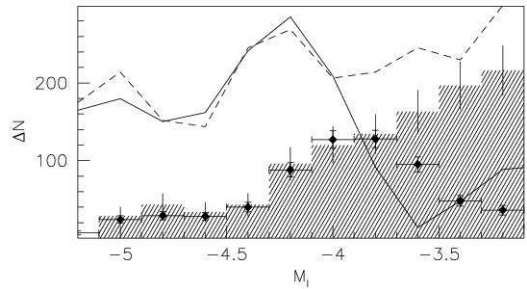}

\caption{
The luminosity function of the Fornax dSph (shaded histogram with
errorbars) is compared to that of the NGC~4038 tail (filled diamonds
with errorbars).  Notice the 
agreement at the bright end,
prior to the point where the Antennae data suffer from incompleteness.
The \protect\( x\protect \) axis represents absolute \protect\( I\protect
\) magnitudes after correcting the two LFs for the respective distances and
foreground absorptions of the two objects.  The 
curves show the results of applying a Sobel filter to the
NGC~4038 (solid curve) and Fornax LFs (dashed curve). Both 
curves show two main peaks, a higher one at $M_I=-4.2$ and a 
smaller one at $M_I \approx -5$. The first one corresponds to 
the RGB tip, while the second one is due to the upper AGB.
\label{fig:fornax-LF}}
\end{figure}

\subsection{The mass in young stars}\label{sec:youngstars}

In order to characterize the star formation and to compare it to what is
observed in dwarf galaxies, we estimate the stellar mass in the young
and old populations.
We start by computing the mass in young stars in the $8$ associations,
based on their main sequence LF.
Since all associations show extended star formation, we will treat star
formation in the tail as a global process, and thus consider an
integrated LF over all associations (see Fig.~\ref{fig:lf-8-log}).
The LF must be complete down to $M_V=-4.2$ (i.e. \( V_{0}=26.5 \)),
since our limiting magnitude is \( V_{0}=28.5 \) and completeness runs
from $1$ to $0$ in $\approx 2$~mag for \noun{wfpc2} imaging in crowding
conditions similar to ours (see e.g. Brocato, Di Carlo \& Menna
\citeyear{brocato_etal01}, and Harris et al. \citeyear{vcc1104}).

A realistic estimation of the mass in young stars needs a model of star
formation, and a simple approach is developed in
Appendix~\ref{sec:simple-sf}.
However, in order to guess the order of magnitude of the mass, and as a
check of our model predictions, we start by computing this quantity
under the hypothesis of a simple stellar population (SSP).
After computing the mass in the $8$ associations, we extrapolate it to
our whole region.  The details of the process are left to
Appendix~\ref{sec:estimating_young}, while here we summarize the main
results.

In order to estimate the 
mass
of a SSP, we can either compare its LF to that of another SSP of known
mass, or use the results of stellar population synthesis. Thus, we first
compare the main sequence LF of the $8$ associations to that of a young
populous cluster of the Large Magellanic Cloud (NGC~2004), and then we
use the theoretical results of Renzini (\citeyear{renzini98}) and
Maraston (\citeyear{maraston98}) to find the size of a young stellar
population whose upper LF matches the observed LF.

In the first case, we find that the $8$ associations contain a factor \(
3.4\pm 1.2 \) more stars than NGC~2004, 
or \( (6.8\pm 2.4)\times 10^{4}\, M_{\odot } \) in young stars (using
the cluster mass given by Mould et al. \citeyear{mould_etal00}).
If we use 
the theoretical
approach, 
then
the resulting mass in young stars is \( (2.1\pm 0.2)\times 10^{4}\,
M_{\odot } \)
(assuming
a mass-to-light ratio
\( M/L_{\rm bol}=0.036 \), and a Salpeter (\citeyear{salpeter})
power-law mass function of exponent \( \alpha =-2.35 \)).  
A few $10^4~M_{\odot}$ of young stars within the eight associations are
then predicted
under the SSP
hypothesis.

\begin{figure}

  \plotone{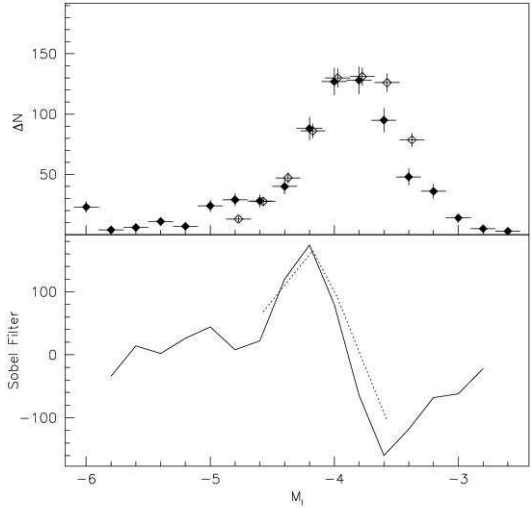}

\caption{
This figure shows that the application of the Sobel filter to our Antennae
data and to a published data set of comparable S/N give the same results.
The $I$ band LF of the faint red stars in the NGC~4038 southern tidal tail
(filled diamonds), and that of the Virgo cluster dwarf elliptical galaxy
VCC1104 (open diamonds), are shown in the upper  panel.
The lower panel shows the LFs after 
convolution with the Sobel edge-detecting filter, 
where the NGC~4038 data are shown with a solid curve, and
the  VCC1104 data are shown as a dotted curve.
 \label{fig:vcc1104}}
\end{figure}

The
inconsistency between the previous mass estimates
could be another sign 
of an extended star formation history.  
Further, the
brighter part of the LF is better fit by a power-law of exponent \(
\alpha =0.48 \) ($0.1$ residual dispersion
in $\log \Delta N$), than by our template SSP power-law of exponent \(
\alpha =0.32 \) ($0.2$ residual dispersion),
and
according to Cole et al.  (\citeyear{cole_etal99}), an exponent
$\alpha=0.48$ is a sign of continuous star formation.  Therefore, in
Appendix~\ref{sec:simple-sf} 
the
mass
of the
young population is computed
under 
this hypothesis.
For this, 
we assume that star formation has been constant up to now, since {\it a
posteriori} one can see that this simplification gives a reasonably good
fit to the observed LF.

The model 
predicts
the past SF rate as
\( \dot{M}=6\pm 3 \times
10^{-3}\, M_{\odot }\, \rm yr^{-1} \), in agreement with the rate computed
by MDL92 from the H\( \alpha \) luminosity of the $3$~H\noun{ii} regions
(and scaled using our distance value). Such rate yields
a mass of
\( 1.1\times 10^{5}\, M_{\odot } \)
produced 
during 
the last $20.6$~Myr
spanned by our LF.
This mass is larger than that computed above, so we see that
the SSP hypothesis leads to an underestimate of the real mass.

By comparing the LF of all MS stars to that of the $8$ associations,
we infer that in the entire region
$\massyoung\,\rm M_\odot$ were converted into
stars since $20.6$~Myr ago.
An absolute constraint on the mass of young stars comes from the total
luminosity measured in this vicinity by MDL92 of $V=17.3$ mag, or $
1.8\times 10^6 L_{V,\odot}$, assuming a distance of $\DMpc$~Mpc.  If
we assume that 100\% of this luminosity is due to young stars, 
it corresponds to $4.5 \times 10^{5} M_\odot$ 
(for the
$M/L$ of a Solar metallicity population of age $100$~Myr and
Salpeter IMF), only a few times larger than our estimate.
So 
the present star formation rate
must
have
been sustained for only a few tens of Myr.
The oldest stars seen in the associations must then have been produced
during a period of lower star formation rate, or in discrete episodes.

\subsection{The mass in old stars} \label{sec:fornax}

As recalled above, stellar populations older than $\sim 2$~Gyr show a
well populated RGB, which ends at an $M_I$ that is almost independent
of age and metallicity (Fig.~\ref{fig:3cmd}). Furthermore, the RGB
luminosity function is a power law whose index is again almost
independent of those two variables. As an example, Saviane et
al. (\citeyear{ivo-tucana}) show that the LF of the dwarf spheroidal
(dSph) in Tucana is similar to that of the Fornax dSph, yet Tucana
hosts a globular cluster-like, metal-poor population while Fornax is a
few Gyr old with an intermediate metallicity.  
Thus, we can estimate the mass of the old population of the tail by
comparing the LF of its RGB with that of another old population of known
mass.
The old population was isolated from the young
populations in \S\ref{sec:cmdsim}, and its CMD was presented in
Fig.~\ref{fig:3cmd}, so we build the corresponding LF 
simply counting stars in magnitude bins.
Unfortunately, the errors of the NGC~4038/9 photometry are so large that
we cannot separate the AGB from the RGB based on their colors, so there
will be some contamination of the LF by stars younger than $\sim
2$~Gyr. On the other hand, the contribution of the AGB to starcounts is
$\sim 2$ orders of magnitude smaller than that of the RGB (see
e.g. Fig.~\ref{fig:match625}), so the mass that we compute will be close
to the correct value.

With respect to an old population with known mass, we
decided to use a dSph galaxy of the Local Group.
These galaxies are ideal 
for this kind of comparison, since they are very close to the Milky Way,
their physical characteristics are known with good accuracy, and they
contain statistically significant numbers of old stars. 
In particular, we use the Fornax dwarf, since it was recently studied in
Saviane et al. (\citeyear{ivo-fornax}), and its photometry is readily
available.  We take the RGB from their field~C (the most populated one)
and, before constructing the LF, we enhance the photometric errors as
was done for our comparison with NGC~625 (\S\ref{sec:cmdsim}). Moreover,
we multiply the counts by a factor $4.8$, since the Fornax field
contains less stars than the NGC~4038/9 field,

For a correct estimation of the mass, we must also remember
that (a)
field C of Saviane et al. contains only $5\%$ of the total population of
Fornax 
(based on a King light profile and parameters from Irwin \& 
Hatzidimitriou \citeyear{irwinHatzi95}); and that (b)
the total area of the \noun{wfpc2} field is
\( 4.1 \) times larger than the portion where the old population was
isolated.

The luminosity functions of the Fornax dwarf and of our NGC~4038/9 field
are compared in Fig.~\ref{fig:fornax-LF}, which shows their close
agreement.  For $-5
\leq M_I \leq -4$, a KS test gives a 
$98.1\%$ probability that the two
distributions share the same parent. 
The Sobel filter applied to the NGC~4038 LF gives the result shown by the
solid line, and the dashed line shows the result for the Fornax
LF. Like in the comparison with NGC~625, we see the coincidence of the
primary peaks at $M_I=-4.2$, corresponding to the RGB tip, and even that
of the secondary peaks at $M_I\approx -5$, corresponding to the upper
AGB. The NGC~4038 curve starts to deviate from that of Fornax below
$M_I=-4$, where incompleteness starts to affect our photometry. 

The Fornax distance modulus and absorption were taken from
Saviane et al. (\citeyear{ivo-fornax}; $(m-M)_0=20.7$, and $A_I=0.058$).
Since that distance was based on the old value for the luminosity of the
RGB tip, a $0.2$~mag correction was added to the distance modulus.
In the comparison with NGC~625, its distance modulus was a free
parameter, while now the Fornax distance is fixed.  The very good match
between the two LFs then means that Fig.~\ref{fig:fornax-LF} not only
can be used to estimate the mass of the old population in the 
NGC~4038 tail, it also provides further strong support to our short
distance.

\begin{figure}

  \plotone{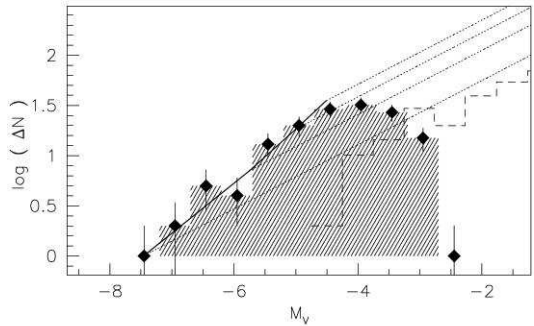}

\caption{
The observed LF of the main sequence can be reproduced (heavy solid
line) with the sum of $3.1$ LFs of simple stellar populations (dotted
lines). We assume that the slope of the single LFs is equal to that of
NGC~2004 (dashed histogram), and that the simple populations contain a
factor $1.7$ more stars than the LMC cluster.  The populations were
generated at a constant rate during $18$~Myr, and all stars now evolved
off the main sequence were removed from the LFs.
\label{fig:lf-sfr}}
\end{figure}

Based on the previous discussion, the number of old stars in the
NGC~4038/9 field, $N_{\rm tail}$, is given
by
\( N_{\rm tail}=N_{\rm Fornax}\times\
4.8\times 4.1\times 0.05 = 0.98\times N_{\rm Fornax}\).  This assumes a
roughly constant density of old stars across the \noun{wfpc2} field.
Since the mass of Fornax is $6.8\times 10^7\, M_{\odot}$ (Mateo
\citeyear{mateo98}), assuming the same $M/L$ we find that there are \(
\massold \, M_{\odot } \) of old stars within the limits of our
\noun{wfpc2} field.  

Again, we can check whether this value is
compatible with the total light measured by MDL92 in their field. If
100\% of the $ 1.8\times 10^6 \, L_{V,\odot}$ is in old stars, then
it would correspond to $\approx 8\times 10^{6}\, M_{\odot}$, if the
Fornax $M/L=4.4$ is used.  The latter mass is smaller than our value,
but our photometry reaches well below the surface brightness limit of
ground-based imaging, so we conclude that our value of the mass for
the old population is a reasonable estimate.

\section{Summary and discussion }\label{sec:discussion}

In this paper we have analyzed the stellar content of a region of
linear dimension $\sim 8$~kpc, in the southern tail of the
Antennae system of interacting galaxies. According to dynamical
simulations (e.g. Barnes 1988), the material of the tail has been
pulled out of NGC~4038 by the tidal field of NGC~4039 $\sim$ 450 
Myr ago (for Milky Way sized progenitors),
and previous groundbased studies of this area have shown that stars
are now actively forming in a few regions (S78, MDL92).  The \noun{wfpc2}
camera on board of the HST has now been used to obtain, for the first
time, broadband $F555W$ and $F814W$ photometry and astrometry of more
than $4000$ stars within the region studied by  MDL92.

A population of faint red stars is seen in the CMD. This population
shows no clumping in its spatial distribution, and it was presumably
ejected from the outer disks of the interacting galaxies. We have
isolated a subsample of this population by selecting only stars from
regions far from the active star forming areas. The selected sample
shows a sharp cutoff in the LF at $I_0<26.5$. The LF is almost
identical to that of the low surface brightness component of the star
forming dwarf irregular galaxy NGC~625, and based on this comparison we
interpreted the cutoff as the RGB tip of a few Gyr old
population. This yields a distance to the Antennae of $\DMpc \pm
\errDMpc$~Mpc, much shorter than the typically quoted $\sim
20$~Mpc. The choice of a shorter distance is reinforced by comparing
the LF of this population to that of VCC1104, a dwarf elliptical
galaxy in the Virgo cluster (see Appendix~\ref{sec:vcc1104}), and to
that of the Fornax dwarf spheroidal galaxy in the Local Group.

Our imaging with \noun{wfpc2} along with the VLA neutral hydrogen map
of Hibbard et al. (\citeyear{hibbard_etal01}) gives a better picture of
the stellar populations at the end of the Southern tail of the
Antennae.  We find that the regions of most active star formation
coincide with the H{\sc i} areas of highest density ($> 8\times
10^{20}~\rm cm^{-2}$), and that the brightest young star forming
associations are similar to those seen in dwarf irregular galaxies. In
particular, there are two elongated structures that contain most of
the associations, and one of them is the ``bar'' identified by
MDL92. Our color-magnitude diagrams confirm the intuitive picture that
the smooth structure of the tail is composed of older stars from the
disks of NGC 4038/9 that were launched along with the gas, due to the
interaction.  The youngest stars are mostly concentrated in OB
associations, but can also be found scattered over the whole region.

In contrast to the inner regions, there are no ``super star clusters''
seen in this region of the tail.  
The young tidal associations are more diffuse and have  orders of
magnitude fewer stars than the central star clusters. 
The eight most prominent
associations have been used to study the recent star formation in the
area. We are able to measure the luminosity function of their main
sequence to reach stars as faint as \( M_{V}=-4.2 \) corresponding
(roughly) to ages of $20$~Myr and masses of \( 14\, M_{\odot } \) (in
the hypothesis of a short distance modulus).  In this range, it would
appear that the star formation has been continuous.  

Considering the stellar mass budget of the whole area that we studied,
by comparing the LF of the red-low surface brightness component to
that of the Fornax dSph galaxy, we find that most of the mass
($\approx 6.7\times 10^7~M_\odot$) is contained in an old ($> 2 $~Gyr)
stellar population, of order the stellar mass of Fornax itself.  With
a simple model of constant star formation we show there is a further
$\approx 2\times 10^5 M_\odot$ of stars formed {\it in situ}
(including young stars outside the eight associations), within the
last $20$~Myr.

Detailed investigations of the star formation process occurring well
outside the boundaries of a disk or the potential well of a galaxy are
rare.  Mould et al. (1999) images the radio jet-induced star formation
near NGC 5128, and finds a population similar to ours, with young OB
associations but no massive star clusters.  Clearly, star formation
requires only the presence of gas of sufficient density, and not the
large scale organizing gravitational potential of a disk, spiral arm,
or galaxy.

One of the most important results of this work is the downward
revision of the distance to the Antennae, which follows the first
determination of the distance modulus based on a standard candle. This
revision has consequences for the the nature of the TDG candidates,
for the stellar populations within the inner regions, for the
dynamical nature of the system, and for the dynamical models of the
local Universe.

With respect to the TDG candidates, Hibbard et al. 
(\citeyear{hibbard_etal01}) used the H{\sc i} kinematics to
explore the dynamical nature of the regions studied by both MDL92 and
S78. Specifically, they evaluate the ratio of the luminous mass, as
inferred from the H{\sc i} and optical luminosity, and the dynamical
mass inferred from the H{\sc i} line width and physical radius. They
derive ratios of $M_{lum}/M_{dyn} \sim 0.3-0.7$ for a distance of 19.2
Mpc, where the range indicates stellar mass-to-light ratios ranging
from $M_*/L_B =$ 0--5 
(the analysis in \S\ref{sec:mass-ssp} suggests
a mass-to-light ratio near the bottom of this range). 
Since the baryonic mass scales as the square of the distance while the
dynamical mass scales linearly with distance, a distance modulus of
\mMo\ requires revising these estimates downward by a factor of 1.4
(i.e. $M_{lum}/M_{dyn} \sim 0.2-0.5$). If either of these regions are
self-gravitating entities, i.e. {\it bona fide} tidal dwarf galaxies,
then most of their mass must be in the form of dark matter. This is
quite hard to understand if they formed from disk material.
Rather than witnessing the formation of a single large dwarf galaxy,
we find it more likely that we are seeing several smaller scale
regions forming stars. The mass scales of these regions are more
typical of dwarf spheroidal galaxies than dwarf irregulars, and they
may have more relevance to remnant streams around galaxies than the
formation of dwarf irregular galaxies (Kroupa 1998).

With respect to the populations believed to result from the merger
induced star formation taking place in the inner regions, all linear
dimensions quoted in W99 would be reduced by a factor of $1.4$, so the
effective radii of the youngest clusters created in the merger would now
be comparable to those of Galactic globulars (W99 find a median
effective radius $\sim 1.5$ times larger than that of GGCs).
The young clusters' LF would of course still be a power law,
however, the bend found by W99 at $M_V\approx -10.4$ would now be a
factor $2$ fainter in luminosity and the corresponding mass would be
smaller, assuming the same $M/L$ value. W99 identified the bend with
the characteristic mass of Milky Way globular clusters, although their
value of $1\times 10^5~M_\odot$ is smaller than that of GGCs ($\approx
2\times 10^5~M_\odot$). With the shorter distance modulus, the mass of
the bend would now be $\approx 5\times 10^4~M_\odot$, making this
identification weaker.

With a smaller distance, the ``Ultra Luminous'' X-ray source
population discovered by Fabbiano, Zezas \& Murray  
(\citeyear{fabbiano_etal2001}; see also Zezas et al. 
\citeyear{zezas_etal2002}, Zezas \& Fabbiano 
\citeyear{zezas_fabbiano2002}), while still extreme in its properties,
becomes both less luminous (peak X-ray luminosity of
$L_X=5\times10^{39}$ ergs s$^{-1}$ vs. $2\times10^{40}$ ergs s$^{-1}$)
and less populous (6 instead of 18 with $L_X>10^{39}$ ergs s$^{-1}$).
The X-ray sources in the Antennae are still luminous, but the new
distance makes their luminosity function comparable to those found in
other starburst galaxies (cf. Zezas \& Fabbiano
\citeyear{zezas_fabbiano2002}).

Finally, with respect to the dynamical models of the local Universe,
we compare the predicted and observed radial velocity of the Antennae
at the new distance, using the Tonry et al. (\citeyear{tonry_etal00})
galaxy flow model of the local Universe, and find that the model
predicts a velocity $500 \kms$ smaller than that of NGC~4038/9. 
The discrepancy 
would hold even if we used the W99 distance (it would then be $>300
\kms$).

Considering the importance of the distance estimate for a physical
understanding of the Antennae, we would urge that the distance modulus
be confirmed with deeper HST/ACS observations.

\begin{figure}

  \plotone{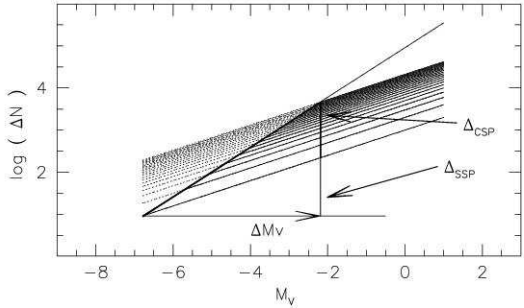} 

\caption{Interpreting the LF of a composite stellar population (heavy
solid line). We assume that it is made by the convolution of several LFs
of simple stellar populations (thin inclined lines). The single
populations are generated at a constant rate $\dot n$, and stars
brighter than the termination of the main sequence are removed from the
LFs (dotted lines).  The figure has been obtained for $\dot{n} =0.21\,
\, \, \rm Myr^{-1} $, from $100$~Myr to $4$~Myr ago.  The labels
identify parameters that are used in the text to develop our model.
\label{fig:sf-labels}
}
\end{figure}

\acknowledgements{}

We thank Brent Tully and St\'ephane Courteau for helpful insight on
galaxy distances in the local supercluster. We also thank Alan Whiting
for making very useful suggestions on the question of the distance to
the Antennae.  
An anonymous referee provided substantial help for the improvement
of an earlier version of this manuscript.
Based on observations made with the NASA/ESA Hubble
Space Telescope, obtained at the Space Telescope Science Institute,
which is operated by the Association of Universities for Research in
Astronomy, Inc. under NASA contract NAS 5-2655. Michael Rich and John
Hibbard acknowledge support from proposal GO-6669, which was provided
by NASA through a grant from the Space Telescope Science
Institute. This research has made use of the NASA/IPAC Extragalactic
Database (NED) which is operated by the Jet Propulsion Laboratory,
California Institute of Technology, under contract with the National
Aeronautics and Space Administration.

\appendix{}

\section{Comparison with a dwarf elliptical galaxy in the Virgo cluster} 
\label{sec:vcc1104}

Harris et al. (\citeyear{vcc1104}) used the HST to image the Virgo
cluster dwarf elliptical galaxy VCC1104 (=IC3388). Photometry in the
$F814W$ filter was obtained with the \noun{wfpc2}, it was calibrated with
groundbased data, and then an LF was created with a bin size 
\( \Delta I=0.05 \). 
If we use the same bin size for the old population of the NGC~4038 tail,
its LF turns out too noisy, since its CMD contains ca. half as many
stars as that of VCC1104. The VCC1104 LF was then rebinned with \(
\Delta I=0.20 \), and the counts were multiplied by $0.43$.
The two LFs are compared in the upper panel of Fig.~\ref{fig:vcc1104},
where absolute magnitudes were computed assuming \( (m-M)_{0}=\mMo \)
for the Antennae, and \( (m-M)_{0}=30.98 \) and \( A_{I}=0.04 \) for the
Virgo dwarf (Harris et al. \citeyear{vcc1104}).  
The lower panel shows the LFs after application of the Sobel filter, and
both curves have a clear peak at \( M_{I}=-4.2 \), although with a large
uncertainty.
The figure shows that the two LFs are strikingly similar for
\( \sim 1 \) magnitude down to \( M_{I}\simeq -3.8 \), where the
incompleteness sets in.  
Hence, the two underlying populations must be similar as well, i.e.
most of the faint red stars of the NGC~4038 tail must be RGB stars.

\section{Estimating the mass of young stars}
\label{sec:estimating_young}

\subsection{Estimates of the mass under the SSP hypothesis} 
\label{sec:mass-ssp}

We start by comparing the brighter part of the LF with that of a young
stellar cluster. Since the template LF must be statistically
significant, we turn to the Large Magellanic Cloud (LMC) clusters. One
of the youngest LMC clusters for which a LF is available is NGC~2004
($16$~Myr, Keller et al. \citeyear{keller_etal00}), and its LF is
plotted in Fig.~\ref{fig:lf-8-log} as open diamonds.  The solid line
through the points represents a fit to the LF, yielding a slope \(
\alpha=0.32\pm 0.04 \).  In the following discussion, this slope will be
used as the template for a simple stellar population. Now since the
two LFs have been formed using the same bins size, they can be
compared directly. We fit a power-law of the same exponent to the
bright part of our LF (as defined above), and the difference in zero
point gives us the scale factor between the NGC~2004 LF and that of
our region. The dotted line in the same figure represents the fit,
which yields a scale factor of \( 3.4\pm 1.2 \).  The error has been
estimated by allowing a \( \pm 1\sigma \) variation in the logarithmic
zero-point. Since the mass of the LMC cluster is \( 2\times 10^{4}\,
M_{\odot } \) (Mould et al. \citeyear{mould_etal00}), this suggests
that our eight associations contain a total of \( (6.8\pm 2.4)\times
10^{4}\, M_{\odot } \) in young stars 

The second way to estimate the 
mass
of a SSP is the following
theoretical one. The number of stars in a given mass range \( m_{1} \)
to \( m_{2} \) on the MS, for a power-law IMF of exponent \( \alpha
\), can be expressed as in Renzini (\citeyear{renzini98}):

\[ N_{\rm MS}=\frac{A}{L_{\rm T}}\, L_{\rm T}\int 
^{m_{2}}_{m_{1}}m^{\alpha }dm\] 

\noindent where $L_{\rm T}$ is the total luminosity of the population, $A$
is a normalization factor, and \( A/L_{\rm T} \) depends on the age of
the population. Assuming that our young stellar population can be
approximated by a single \( \simeq 30\, \rm Myr \) old burst (the
youngest age available in the adopted models), then the computations
of Maraston (\citeyear{maraston98}) give \( A/L_{\rm T}\simeq 0.007 \)
and a mass to light ratio \( M/L_{\rm bol}=0.036 \). Now we count \(
N_{\rm MS}=74 \pm 9 \) stars from \( M_{V}=-7.7 \) down to \(
M_{V}=-4.2 \), and using the Girardi et al. (\citeyear{girardi00})
isochrones (for \( Z=0.008 \)), these luminosities can be translated
into a mass range from \( m_{1}=14 \, M_{\odot } \) to \( m_{2}=55\,
M_{\odot } \).  Introducing these numbers into the equation, for a
Salpeter (\citeyear{salpeter}) \( \alpha =-2.35 \) we obtain \( L_{\rm
T}=(5.9\pm 0.7)\times 10^{5}\, L_{\odot } \), or \( M_{\rm T}=(2.1\pm
0.2)\times 10^{4}\, M_{\odot } \) using the quoted $M/L$.

\subsection{Estimate of the mass under the extended star formation
 hypothesis} 

For this estimate, we use the model developed in the next section, which
allows to compute the number of SSPs created per unit time ($\dot{n}_{\rm
SSP}$) during the recent SF history, as a function of  
the slope of the composite LF $\alpha_{\rm CSP}$, the magnitude interval
it spans  $\Delta M_V$, and the corresponding time interval.
In particular we have $\alpha_{\rm CSP}=0.49\pm 0.05$ (for $23.5 \leq V_0\leq
26.5$),  $\Delta M_V=3$~mag, and a time interval from $20.6$
to $2.5$~Myr ago.
Inserting these values in Eq.~(\ref{e:sfr}), we find a SSP formation
rate of $\dot{n} = 1.7 \pm 0.8 \times 10^{-7}$~SSPs~yr$^{-1}$.  Our
model LF (heavy solid line) is plotted over the LF of the $8$
associations in Fig.~\ref{fig:lf-sfr}.  The model predicts that $3.1$
SSPs were created in \( \approx 18 \)~Myr. Of course, we have already 
shown that stars older than $20.6$~Myr (and as old as, at least, 
$\sim 100$~Myr) are also present in the associations, but due to the 
incompleteness of our photometry, we cannot use the main sequence LF 
to probe the older star formation. In Fig.~\ref{fig:lf-sfr}, the dotted 
line that reaches $M_V \approx -7.5$ represents the LF of one of the 
simple stellar populations. Comparing it to that of NGC~2004 (dashed 
histogram), we find that one generation produces \( 3.4\times 10^{4}\, 
M_{\odot } \), so in terms of mass, the past star formation rate
was \( \dot{M}=6\pm 3 \times 10^{-3}\, M_{\odot }\, \rm yr^{-1} \),
and
we can finally estimate that \( 1.1\times 10^{5}\, M_{\odot } \)
have been produced within the $8$ associations in the last $20.6$~Myr.

We can compare 
our SF
rate to that inferred from the H\( \alpha \)
luminosity of the $3$~H\noun{ii} regions measured by MDL92.
The total luminosity is \( \sum L(H\alpha )= 3.9\times 10^{39} 
\)~erg~s\( ^{-1} \), for their distance of $33.2$~Mpc. Using the 
conversion of Hunter et al. (\citeyear{hunter_etal86})\footnote{
\( \dot{M}=7.07\times 10^{-42}\eta ^{-1}L_{\rm H\alpha }\, \, \, 
M_{\odot }\rm yr^{-1} \) (\( \eta \simeq 0.5 \))}, and adopting our
revised distance of $\DMpc \pm \errDMpc$~Mpc yields a star formation
rate of \( \dot{M}=9.5\pm 2.5\times 10^{-3}\, M_{\odot }\, \rm yr^{-1} 
\), which is in agreement, within the errors, with our derivation for 
the $8$ associations. Note also that a value $\eta=0.8$ in the Hunter
et al. formula, would reconcile the star formation rate from the
H$\alpha$ luminosity to our finding.  The agreement between the two
values is not unexpected, since most of the young stars are contained
in the $3$ associations studied by MDL92.

\subsection{A simple model of star formation }\label{sec:simple-sf}

The purpose of our model is to explain the observed shape of the main
sequence LF, in the case of continuous and constant star formation.
The model is illustrated by
Fig.~\ref{fig:sf-labels}, where we assume that 
such
star formation can be approximated by the sum of a discrete number of
simple stellar populations (SSPs).  Each generation of stars produces a
power-law LF which is added on top of that of the previous generation
(thin solid inclined lines). As time passes, fainter and fainter stars
die, so we truncate the single LFs at fainter and fainter magnitudes,
with truncation magnitude as a function of time taken from the
theoretical isochrones of Girardi et al.  (\citeyear{girardi00}). The
observed LF (thick solid line) will then be the convolution of the LFs
of several generations of stars, where the number of generations depends
on the star formation efficiency and duration. The slope of the observed
LF will be equal to that of a single generation of stars for magnitudes
fainter than the current termination of the MS (TMS), while the slope
will be larger for magnitudes brighter than the TMS. In this latter
range, the slope will depend on the number of generations per unit time
and on the slope of the single LF.

The previous reasoning can be formalized as follows. We assign
variables $\alpha_{\rm SSP}$ and $b_{\rm SSP}$ as the slope and
intercept of the single LF in the $M_V,\Delta N$ plane shown in
Fig.~\ref{fig:sf-labels}.  Then, let's express the star formation rate
as the number of SSPs created in the unit time, \( \dot{n}_{\rm SSP}
\).  In a given interval \( \Delta t \), a number of populations \(
N_{\rm SSP}=\dot{n}_{\rm SSP}\times \Delta t \) is then created. In the
absence of stellar deaths, the final LF would be a power-law of the same
initial slope, and the zero point of this composite stellar population
would be \( b_{\rm CSP}= b_{\rm SSP}+\log (N_{\rm SSP}) \).  Instead,
each LF we add will have a fainter TMS for increasing age, so each LF in
the figure has been truncated at the corresponding \( M^{\rm TMS}_{V} \)
for its age.  The age is
\( t=t_{\rm min}+i\times T_{1} \), for the \( i^{\rm th} \)
population, where \( T_{1}=1/\dot{n}_{\rm SSP} \).
Now from Fig.~\ref{fig:sf-labels}, we see that the slope of the brightest
portion of the LF is \( \alpha _{\rm CSP}=(\Delta_{\rm
CSP}+\Delta_{\rm SSP})/\Delta M_{V} \), where $\Delta_{\rm CSP}$,
$\Delta_{\rm SSP}$, and $\Delta M_{V}$ are defined in the figure. If
we assume that the LFs of all generations have the same slope, then
all quantities in the equation can be expressed as functions of that
slope, the star formation rate, and the times of the beginning and end
of the star formation. Thus, from the equation we can compute the star
formation rate.

The horizontal offset is set by the difference between the magnitude of
the TMS at the oldest and youngest ages. We found the time dependence of
the TMS on the basis of the Girardi et al. (\citeyear{girardi00})
isochrones. After plotting \( M^{\rm TMS}_{V} \) vs. \( \log t \) for
several combinations of metallicity and age intervals, we found that a
good approximation is
\begin{equation}\label{e:mv-tms} 
M^{\rm TMS}_{V}=3.29\times \log t-28.51
\end{equation} 
which is valid in the interval \( 6.6\leq \log t\leq 9.5 \), and for 
metallicities \( 0.004\leq Z\leq 0.02 \). The $1$-$\sigma$ error on the 
coefficients is \( 1\% \), and the rms dispersion of the data around the
fit is \( 0.4 \) magnitudes. 
From  expression (\ref{e:mv-tms}), it follows that
\( \Delta M_{V}=3.29\times \log (t_{\rm max}/t_{\rm min}) \).
As it has been shown above, the vertical offset introduced by creating
\( N_{\rm SSP} \) populations, is \( \Delta_{\rm CSP}=b_{\rm CSP}-b_{\rm
SSP}=\log N_{\rm SSP} \), and since \( N_{\rm SSP}=\dot{n}_{\rm
SSP}\times \Delta t \), it is \( \Delta_{\rm CSP}=\log [\dot{n}_{\rm
SSP}\times (t_{\rm max}-t_{\rm min})] \).  Moreover, the vertical
offset due to the simple stellar population LF is of course \( \Delta_{\rm
SSP}=\alpha _{\rm SSP}\times \Delta M_{V}=\alpha _{\rm
SSP}\times 3.29 \times \log (t_{\rm max}/t_{\rm min}) \), where
$\alpha_{\rm SSP}$ is the slope of the SSP LF.
In conclusion, the expected slope of a composite population that has
witnessed continuous star formation from \( t_{\rm min} \) to \( 
t_{\rm max} \), at a rate \( \dot{n}_{\rm SSP} \), is
\[ 
\alpha _{\rm CSP}=\alpha _{\rm
SSP}+\frac{\Delta_{\rm CSP}}{\Delta M_{V}}=\alpha _{\rm
SSP}+\frac{\log [\dot{n}_{\rm SSP}\, (t_{\rm max}-t_{\rm
min})]}{3.29\, \log (t_{\rm max}/t_{\rm min})}
\] 
From this
expression we can now obtain the number star formation rate as a function 
of the observables 
\begin{equation}\label{e:sfr} 
\log \dot{n}_{\rm SSP}=(\alpha _{\rm
CSP}-\alpha _{\rm SSP})\, \Delta M_{V}-\log \Delta t
\end{equation}
where an estimate of \( \log \Delta t \) can be obtained from
Eq.~(\ref{e:mv-tms}) as a function of the observed range in \( M_{V}
\) along the LF.


\begin{thebibliography}{}
\bibitem[(1999)]{allen} Cox, A.N. (ed), 1999, {\it Allen's Astrophysical
Quantities}, Fourth Edition, Los Alamos, NM
\bibitem[(1992)]{barnes-hernquist92}Barnes, J.E., Hernquist, L. 1992,
Nature, 360, 715
275
\bibitem[(2001)]{brocato_etal01} Brocato, E., Di Carlo, E., Menna,
G.  2001, A\&A, 374, 523
\bibitem[(2000)]{carretta_etal00} Carretta, E., Gratton, R.~G., Clementini,
G., \& Fusi Pecci, F.\ 2000, ApJ, 533, 215
\bibitem[(1999)]{cole_etal99}Cole, A.A., et al. 1999, AJ, 118, 1657
\bibitem[(1998)]{dacosta-canarie}Da Costa, G. S. 1998, \emph{Stellar astrophysics for the local group}: VIII
Canary Islands Winter School of Astrophysics, 351
\bibitem[(1958)]{supercluster58}de Vaucouleurs, G. 1958, AJ, 63, 253
\bibitem[(1968)]{ngc-4027}de Vaucouleurs, G., de Vaucouleurs, A., Freeman, K.B. 1968, MNRAS, 139, 425
\bibitem[(2000)]{dolphin-cte}Dolphin, A.E. 2000, PASP, 112, 1397 (D00)
\bibitem[(1998)]{duc_mirabel98}Duc, P.-A., Mirabel, I.F. 1998, A\&A, 333, 813
\bibitem[(1997)]{duc_etal97}Duc, P.-A., Brinks, E., Wink, J.E., Mirabel, I.F. 1997, A\&A, 326, 537
\bibitem[(1997)]{elmegreen97}Elmegreen, B. G., \& Efremov, Y. N. 1997, ApJ, 480, 235
\bibitem[(2001)]{fabbiano_etal2001}Fabbiano, G., Zezas, A., \& Murray, S.S. 2001 ApJ, 554, 1035
\bibitem[(2000)]{laura-m87}Ferrarese L., et al. 2000, ApJ, 529, 745
\bibitem[(2000)]{girardi00}Girardi, L., Bressan, A., Bertelli, G., Chiosi, C. 2000, A\&AS, 141, 371.
\bibitem[(2001)]{gordon_etal01}Gordon, S., Koribalski, B., \& Jones, K.\ 2001, MNRAS, 326, 578
\bibitem[(1998)]{vcc1104}Harris, W.E., Durrell, P.R., Pierce, M.J., Secker, J. 1998, Nature, 395, 45
\bibitem[(2001)]{hibbard_etal01}Hibbard, J.~E., van der Hulst, J.~M., Barnes, J.~E., \& Rich, R.~M.\ 2001, AJ, 122, 2969
\bibitem[(1995)]{holtzman}Holtzman, J.A., et al. 1995, PASP, 107, 1065 (H95)
\bibitem[(1986)]{hunter_etal86}Hunter, D.A., Gillett, F.C., Gallagher, J.S., Rice, W.L., Low, F.J. 1986, ApJ,
303, 171
\bibitem[(1999)]{ibata-m4}Ibata, R.A. 1999, ApJS, 120, 265 (I99)
\bibitem[(1995)]{irwinHatzi95}Irwin, M. \& Hatzidimitriou, D., 1995,
MNRAS, 277, 1354 
\bibitem[(1998)]{jerjen_etal98} Jerjen, H., Freeman, K. C., \& Binggeli,
B. 1998, AJ, 116, 2873
\bibitem[(1992)]{jogDas92}Jog, C. J., Das, M. 1992, ApJ, 400, 476
\bibitem[(1996)]{jogDas96}Jog, C. J., Das, M. 1996, ApJ, 473, 797
\bibitem[Kauffmann \& White(1993)]{kauffmannWhite93} Kauffmann, G.~\& White, S.~D.~M.\ 1993, \mnras, 261, 921 
\bibitem[(2000)]{keller_etal00}Keller, S.C., Bessell, M.S., Da Costa, G.S. 2000, AJ, 119, 1748
\bibitem[(1993)]{lfm93}Lee, M.G., Freedman, W.L., Madore, B. F. 1993,
ApJ, 417, 553
\bibitem[(1996)]{lineweaver_etal96} Lineweaver, C. H., Tenorio, L.,
Smoot, G. F., Keegstra, P., Banday, A. J., \& Lubin, P. 1996, ApJ, 470, 38
\bibitem[(1995)]{bf95}Madore, B.F., Freedman W.L. 1995, AJ, 109, 1645 (MF95)
\bibitem[(1998)]{maraston98}Maraston, C., 1998, MNRAS, 300, 872
\bibitem[(1998)]{mateo98}Mateo, M., 1998, ARA\&A, 36, 435
\bibitem[Mirabel et al. (1992)]{mirabel}Mirabel, I.F., Dottori, H., Lutz, D. 1992, A\&A, 256, L19 (MDL92)
\bibitem[(2000)]{mould_etal00} Mould, J. R., et al., 2000, ApJ, 536, 266
\bibitem[(2000)]{nikolaev_weinberg00} Nikolaev, S.,  \&  Weinberg,
M. D., 2000, ApJ, 542, 804
\bibitem[(1985)]{dV-4027}Pence, W.D., de Vaucouleurs, G. 1985, ApJ, 298,
560
\bibitem[(2002)]{piotto_etal02} Piotto, G., et al. 2002, A\&A, 391, 945
\bibitem[(1998)]{renzini98}Renzini, A., 1998, AJ, 115, 2459
\bibitem[(1999)]{rozas_etal99} Rozas, M., Zurita, A., Heller, C.H., \&
Beckman, J.E., 1999, A\&AS, 135, 145
\bibitem[(1970)]{rubin}Rubin, V.C., Ford, W.K., D'Odorico S. 1970, ApJ,
160, 801
\bibitem[(1955)]{salpeter} Salpeter, E. E. 1955, ApJ, 121, 161
\bibitem[(1996)]{ivo-tucana} Saviane, I., Held, E.V., Piotto, G. 1996,
A\&A, 315, 40
\bibitem[(2000)]{ivo-fornax}Saviane, I., Held, E.V., Bertelli, G. 2000, A\&A, 355, 56
\bibitem[(2003)]{saviane_etal02} Saviane, I., Rosenberg, A., Piotto G., \& Aparicio,
A. 2003. In
New Horizons in Globular Cluster Astronomy, ed. G. Piotto,
G. Meylan, G. Djorgovski, and M. Riello, ASP Conf. Ser., 296, p. 402
\bibitem[(1998)]{schlegel_etal98}Schlegel, D. J., Finkbeiner, D. P. \& Davis, M., 1998, ApJ, 500, 525 
\bibitem[Schweizer (1978)]{schweizer78}Schweizer, F. 1978, in Structure and Properties of Nearby Galaxies, eds. E.
M. Berkhuijsen and R. Wielebinski (Dordrecht,Reidel), p. 279 (S78)
\bibitem[(1987)]{stetson87}Stetson, P.B. 1987, PASP, 99, 191
\bibitem[(1994)]{stetson94}Stetson, P.B. 1994, PASP, 106, 250
\bibitem[(2000)]{tonry_etal00} Tonry, J.L., Blakeslee, J.P., Ajhar,
E.A., \& Dressler, A., 2000, ApJ, 530, 625
\bibitem[(2001)]{tonry4033} Tonry, J.~L., Dressler,
 A., Blakeslee, J.~P., Ajhar, E.~A., Fletcher, A., Luppino, G.~A.,
 Metzger, M.~R., \& Moore, C.~B.\ 2001, \apj, 546, 681
\bibitem[(1972)]{TT}Toomre, A., \& Toomre, J. 1972, ApJ, 178, 623
\bibitem[(1988)]{TullyCatalog} Tully, B. 1988, {\it Nearby Galaxies Catalog},
Cambridge University Press
\bibitem[(1996)]{vacca_etal96} Vacca, W.D., Garmany, C.D., Shull, J.M.,
1996, ApJ, 460, 914
\bibitem[(1999)]{whitmore}Whitmore, B.C., Zhang, Q., Leitherer, C., Fall, S. M., Schweizer, F., Miller,
B.W. 1999, AJ, 118, 1551
\bibitem[(2000)]{wilson00}Wilson C.D., Scoville N., Madden S.C., Charmandaris V. 2000, ApJ, 542, 120
\bibitem[(2002)]{zezas_etal2002}Zezas, A., Fabbiano, G., Rots, A.H.,
\& Murray, S.S. 2002, ApJS, 142, 239
\bibitem[(2002)]{zezas_fabbiano2002}Zezas, A., \& Fabbiano, G. 2002,
ApJ, 577, 726
\bibitem[Zwicky (1956)]{zwicky56}Zwicky, F. 1956, Ergebnisse der Exakten Naturwissenschaften, 29, 344
\end{thebibliography}
\end{document}